\documentclass[aps, prx, twocolumn, superscriptaddress]{revtex4-2}
\usepackage{graphicx}
\usepackage{tikz}
\usepackage{dcolumn}
\usepackage{times}
\usepackage[braket, qm]{qcircuit}
\usepackage{physics}
\usepackage{mathtools}
\usepackage{wasysym}
\usepackage{amsmath}
\usepackage{amssymb}
\usepackage{amsfonts}
\usepackage{hhline}
\usepackage{multirow}
\usepackage{hhline}
\usepackage{xcolor}
\usepackage[final]{listings}
\usepackage{hyperref}
\usepackage{graphicx}

\definecolor{darkgreen}{rgb}{0,0.5,0}
\definecolor{darkred}{rgb}{0.5,0,0}
\definecolor{darkblue}{rgb}{0,0,0.5}
\definecolor{MediumGray}{gray}{0.60}

\lstdefinestyle{pseudo}{
	language=Python,
	xleftmargin=5.0ex,
	moredelim = [s][\textit]{[}{]},
	basicstyle={\small\ttfamily},
	captionpos=b,
	columns=flexible,
	numbers=left,
	numberstyle=\small,
	numbersep=8pt,
	stepnumber=1,
	numberstyle=\tiny\color{gray},
	escapechar=\%,
	breaklines=true,
	frame=single,
	framexleftmargin=15pt,
	tabsize=2,
	postbreak=\mbox{\textcolor{red}{$\hookrightarrow$}\space},
	captionpos=b,
	showspaces=false,
	showtabs=false,
	keywords=[3]{QuantumState, List, Tuple},
	commentstyle=\itshape\color{MediumGray},
}

\begin{document}

\title{Implementing Fault-tolerant Entangling Gates on the Five-qubit Code and the Color Code}

\author{C. Ryan-Anderson}
\email{ciaran.ryan-anderson@quantinuum.com}
\author{N. C. Brown}
\email{natalie.brown@quantinuum.com}
\author{M. S. Allman}
\author{B. Arkin}
\affiliation{Quantinuum, 303 South Technology Ct., Broomfield, CO 80021, USA}
\author{G. Asa-Attuah}
\affiliation{Quantinuum, 1985 Douglas Dr. N,
Golden Valley, MN 55422, USA}
\author{C. Baldwin}
\author{J. Berg}
\author{J. G. Bohnet}
\author{S. Braxton}
\affiliation{Quantinuum, 303 South Technology Ct., Broomfield, CO 80021, USA}
\author{N. Burdick}
\affiliation{Quantinuum, 1985 Douglas Dr. N,
Golden Valley, MN 55422, USA}
\author{J. P. Campora}
\author{A. Chernoguzov}
\author{J. Esposito}
\author{B. Evans}
\author{D. Francois}
\author{J. P. Gaebler}
\author{T. M. Gatterman}
\author{J. Gerber}
\author{K. Gilmore}
\author{D. Gresh}
\author{A. Hall}
\author{A. Hankin} 
\affiliation{Quantinuum, 303 South Technology Ct., Broomfield, CO 80021, USA}
\author{J. Hostetter}
\affiliation{Quantinuum, 1985 Douglas Dr. N,
Golden Valley, MN 55422, USA}
\author{D. Lucchetti}
\author{K. Mayer}
\affiliation{Quantinuum, 303 South Technology Ct., Broomfield, CO 80021, USA}
\author{J. Myers}
\affiliation{Quantinuum, 1985 Douglas Dr. N,
Golden Valley, MN 55422, USA}
\author{B. Neyenhuis}
\affiliation{Quantinuum, 303 South Technology Ct., Broomfield, CO 80021, USA}
\author{J. Santiago}
\author{J. Sedlacek}
\affiliation{Quantinuum, 1985 Douglas Dr. N,
Golden Valley, MN 55422, USA}
\author{T. Skripka}
\affiliation{Quantinuum, 303 South Technology Ct., Broomfield, CO 80021, USA}
\author{A. Slattery}
\affiliation{Quantinuum, 1985 Douglas Dr. N,
Golden Valley, MN 55422, USA}
\author{R. P. Stutz}
\affiliation{Quantinuum, 303 South Technology Ct., Broomfield, CO 80021, USA}
\author{J. Tait}
\affiliation{Quantinuum, 1985 Douglas Dr. N,
Golden Valley, MN 55422, USA}
\author{R. Tobey}
\affiliation{Quantinuum, 303 South Technology Ct., Broomfield, CO 80021, USA}
\author{G. Vittorini}
\affiliation{Quantinuum, 1985 Douglas Dr. N,
Golden Valley, MN 55422, USA}
\author{J. Walker}
\author{D. Hayes}
\affiliation{Quantinuum, 303 South Technology Ct., Broomfield, CO 80021, USA}

\date{\today}

\begin{abstract}
We characterize and compare two different implementations of fault-tolerant entangling gates on logical qubits. In one instance, a twelve-qubit trapped-ion quantum computer is used to implement a non-transversal logical CNOT gate between two logical qubits using the $[[5,1,3]]$ quantum error correction code. The operation is evaluated with varying degrees of fault tolerance, which are provided by including quantum error correction circuit primitives known as flagging and pieceable fault tolerance. In the second instance, a twenty-qubit trapped-ion quantum computer is used to implement a transversal logical CNOT gate on two logical qubits using the $[[7,1,3]]$ color code. The two codes were implemented on different but similar devices, and in both instances, all of the quantum error correction primitives, including the determination of corrections via decoding, are implemented during runtime using a classical compute environment that is tightly integrated with the quantum processor. For different combinations of the primitives, logical state fidelity measurements are made after applying the gate to different input states, providing bounds on the process fidelity. We find the highest fidelity operations with the $[[7,1,3]]$ color code, with the fault-tolerant state preparation and measurement (SPAM) operation achieving fidelities of $0.99939(15)$ and $0.99959(13)$ when preparing eigenstates of the logical $\overline{X}$ and $\overline{Z}$ operators, which is higher than the average physical qubit SPAM fidelities of $0.9968(2)$ and $0.9970(1)$ for the physical $X$ and $Z$ bases, respectively. When combined with a logical transversal CNOT gate, we find the $[[7,1,3]]$ color code to perform the sequence ~\textendash~ state preparation, CNOT, measure out ~\textendash~ with an average fidelity bounded by $[0.9957,0.9963]$. The logical fidelity bounds are higher than the analogous physical-level fidelity bounds, which we find to be $[0.9850,0.9903]$, reflecting multiple physical noise sources such as SPAM errors for two qubits, several single-qubit gates, a two-qubit gate and some amount of memory error. To help assess the long-term promise of these codes, we also present detailed simulations of the two codes' performance in regimes of lower physical error rates.
\end{abstract}
\maketitle
\thispagestyle{empty} 

It is widely believed that quantum error correction (QEC) will be necessary to achieve the error rates required for large computations such as Shor's factoring algorithm~\cite{shor1999polynomial}, and researchers have begun exploring the real-world operation of QEC codes~\cite{hilder2021faulttolerant,Linkee1701074,PhysRevLett.119.180501,harper2019fault,gong2021experimental,erhard2020entangling,egan2020fault,chen2021exponential,chiaverini2004realization,schindler2011experimental,riste2020real,heeres2017implementing,campagne2020quantum,RyanAnderson2021,Gong2021,Abobeih2021,Postler2021}. Much progress has been made in the theoretical understanding of fault tolerance (FT)~\cite{Gottesman1998,Eastin09,reichardt2020fault}, but it is not yet clear which innovations will prove the most enabling for building universal quantum computers. Different codes and protocols come with a variety of advantages and disadvantages with respect to encoding rates, error thresholds, parity check circuit complexity and geometry, natural gate sets, and FT circuitry overhead, and these requirements will have to meet real-world technical constraints before clear advantages can emerge.

FT circuit design aims to suppress the error rates of arbitrary quantum algorithms. How to define and apply FT is not universally agreed upon, but here we take FT to refer to circuit design principles that guarantee correctable faults do not spread too quickly through circuits to become uncorrectable logical errors~\cite{aliferis2006exrec,gottesman2010introduction}. QEC codes are able to correct a finite number of errors, so if a possibly corrupted qubit interacts with too many other qubits before the code attempts to make corrections, the protocol may not be FT and result in high logical error probability. Since qubits must interact with one another, it is inevitable that faults spread within circuits; however, it is possible to examine every possible origin of the most likely faults and trace them through a circuit to determine if a QEC protocol (including logical operations) can limit the spread of faults and correct them up to the corrective power of the code. While FT and noise suppression are thought to be crucial to large scale quantum computing, that does not imply the need to correct \textit{every} possible error, (\textit{e.g.}, all qubits disappear). Indeed, QEC codes are designed to correct only a subset of all possible errors, set by the \textit{code distance}, keeping the design problem tractable.

Typically, QEC uses ancilla qubits to make parity (syndrome) measurements of data qubits~\cite{Nielsen00}, a process known as syndrome extraction. There are different schemes for syndrome extraction (\textit{e.g.}, Shor \cite{shor1996fault}, Steane \cite{steane1997active}, and Knill \cite{knill2005quantum} style), but one of the most qubit-efficient methods uses only a single ancilla qubit (``bare ancilla") per parity measurement. However, even when designed with care, some syndrome measurement circuits utilizing bare ancillas can cause faults to spread ruinously through the circuit, leading to logical errors that would otherwise be correctable if caught earlier. These types of errors are known as hook errors \cite{dennis2002topological, RyanAnderson2021}.  For some codes, such as the surface code, syndrome measurements using bare ancillas can be specially scheduled to avoid hook errors~\cite{tomita2014low}; however, other codes require additional techniques. One such technique to protect against hook errors when using bare ancillas is known as the flagging scheme~\cite{Reichardt18,yoder2017surface}, a scheme that we use extensively in this work.

In addition to syndrome extraction circuits, computational circuits (logical gates) must also be FT. The most straightforward circuit structure for implementing a FT logical gate is a transversal gate ~\cite{Nielsen00}, meaning the logical gate can be implemented by applying a single physical gate to each qubit. In the case of logical two-qubit gates, a transversal two-qubit gate is often implemented by using physical two-qubit gates that connect corresponding physical qubits in the different code blocks. However, QEC codes based on \textit{qubits} admit only a limited and non-universal set of transversal gates~\cite{Eastin09}. For example, the $[[5,1,3]]$ five-qubit code (or the perfect code)~\cite{Laflamme96} admits a full set of single-qubit Clifford transversal gates but not a transversal CNOT when restricted to two logical qubits~\cite{Gottesman1998}. 

For the price of two additional physical data qubits per logical qubit, the $[[7,1,3]]$ color code (or the Steane code)~\cite{steane1996error} not only admits the full set of single-qubit Clifford transversal gates, but also a transversal multi-qubit Clifford gate. As we discuss at length in this work, the transversal two-qubit gate offers significant savings in computational circuitry overhead, thus highlighting an important trade in the design of FT quantum computing architectures: qubit number vs circuit depth. Note, for brevity, in this paper we will simply refer to the $[[5,1,3]]$ five-qubit code simply as the five-qubit code and the $[[7,1,3]]$ code simply as the color code.

\begin{figure}[ht]
\includegraphics[trim=0 0 0 0, clip,width=\columnwidth]{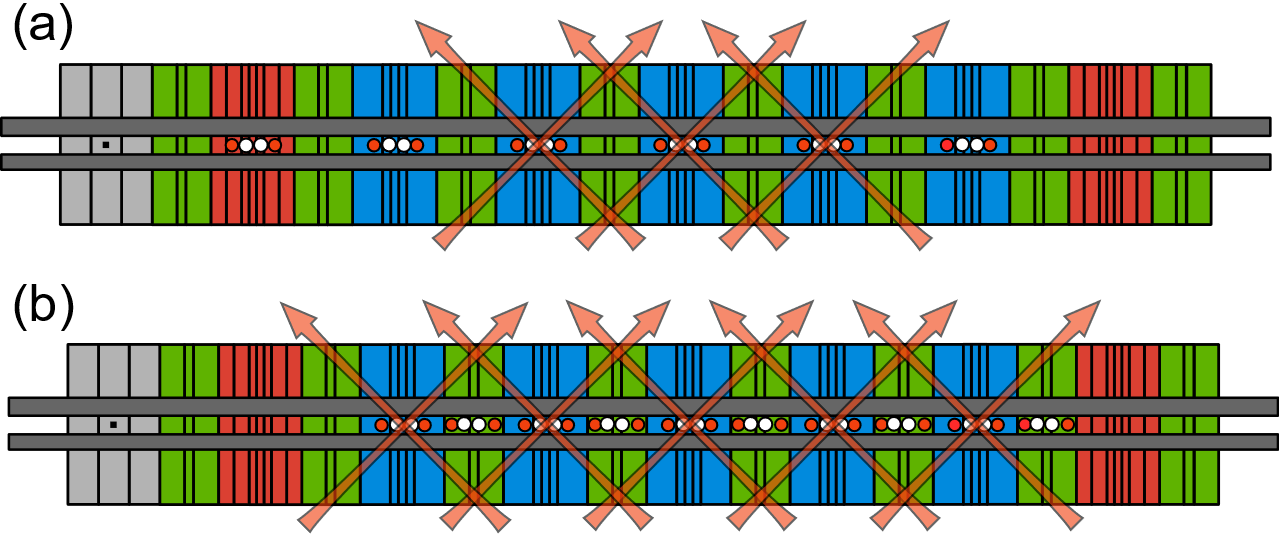}
\caption{An illustration of the (a) H1-2 and (b) H1-1 systems as they were configured for the five-qubit code and color code experiments. H1-2 has three zones capable of performing parallel gates, marked by the crossing laser beams, and the system can use up to twelve qubits. H1-1 has five parallel gate zones, and the system can use up to twenty qubits.The $^{171}\textrm{Yb}+$ qubit ions (red circles) are always paired with $^{138}\textrm{Ba}+$ ions (white circles) for sympathetic cooling. Note that the ion crystals are not drawn to scale as the gating zones are $750 \mu$m apart and the four-ion crystals are $8 \mu$m long.}
\label{fig_n12_and_n20_traps}
\end{figure}

In this work, we begin to explore the trades in implementing FT logical operations on QEC codes in real devices. We use two different quantum computers in this trade study, both based on the same QCCD~\cite{Wineland98} design described in Ref.~\cite{Pino2020}, but with different limitations. The first system used (Fig.~\ref{fig_n12_and_n20_traps} (a)) is configured with twelve qubit ions and is used to implement a logical CNOT gate between two five-qubit code encoded qubits. 

The second system (Fig.~\ref{fig_n12_and_n20_traps}(b)) is configured with twenty qubit ions and is able to support a logical CNOT gate between two color code qubits~\cite{steane1996error,Postler2021}. While entangling operations between logical qubits have previously been demonstrated~\cite{Postler2021}, this work represents the first demonstration of a logical entangling operation carried out fault-tolerantly with real-time QEC being embedded. We assess the quality of the operations by evaluating the state fidelity for different input states, allowing us to place bounds on the process fidelity (a.k.a. entanglement fidelity, see Appendix~\ref{appendix_fidelity}).

The experiments on the different codes were done using different machines with slightly different noise environments. To aid with comparisons in this trade study, we supplement our measurements with detailed simulations to evaluate the QEC protocols under the same noise environments. In general, we find that the color code has higher performance and that the five-qubit code would likely only be advantageous in systems with physical error rates far below what can be achieved today.

\section{The quantum computers}

This work reports results from running two different codes on two different quantum computers: the five-qubit code and the color code running on the Quantinuum H1-2 and H1-1 systems, respectively. The two H1 series systems are similar in that both use the same surface electrode ion trap to control $^{171}\text{Yb}^+$ ions as qubits in the QCCD quantum processor architecture. This architecture enables a fully reconfigurable qubit register, enabling all-to-all qubit connectivity via ion transport operations~\cite{Kaushal20, Pino2020}.  Ion transport to isolated gate zones with focused laser beams (Fig.~\ref{fig_n12_and_n20_traps}) also provides low crosstalk gate and mid-circuit measurement operations, crucial for FT QEC where error syndromes are measured during computation. Each qubit ion is accompanied by a $^{138}\text{Ba}+$ ion used for sympathetic cooling~\cite{Barrett03} during transport and immediately before two-qubit gate operations. Gates can be executed simultaneously in dedicated gate zones using off-resonant stimulated Raman transitions that are characterized using randomized benchmarking~\cite{Emerson2005,KnillRB}. Likewise, SPAM operations are done in parallel in each gate zone. The typical error rates for these systems are in Table~\ref{error_table}.

While the two systems are overall quite similar, H1-1 has some system upgrades that enable experiments with two logical qubits encoded with the color code. H1-2 implements two logical five-qubit code qubits using twelve physical qubits with three active gate zones.  H1-1 can use up to twenty physical qubits in five active gate zones. Additionally, H1-1 implements a routine to correct phase errors arising from qubit frequency differences across the ion trap. We refer to this routine as spatial phase tracking. Spatial inhomogeneities in the magnetic field~\textendash~the magnitude of which are discussed later in section \ref{sec_simulations}~\textendash~create position-dependent qubit frequencies.  This leads to relative dephasing or phase errors between the qubits. The spatial phase tracking routine uses the measured spatial qubit frequency information along with the ion transport compiler to generate adjustments to laser pulses driving the physical gate operations. Correcting the spatial phase errors amounts to adding small $Z$ rotations to single-qubit operations. 

In previous experiments~\cite{RyanAnderson2021}, we implemented real-time decision making using an extended version of OpenQASM 2.0~\cite{cross2017open} that supports elementary feed-forward operations conditioned on measurement results. Since then, we have integrated a more capable classical compute resource. Compared to our previous methods, the new classical co-processor has significantly enhanced capabilities which will be essential to scalable algorithmic decoders~\cite{dennis2002topological}. For this work, we used the classical co-processor to implement look-up table decoders.

\begin{table}[ht]
\resizebox{\columnwidth}{!}{%
\begin{tabular}{|l|c|c|c|c|}
\hline
System & SQ & TQ & SPAM & MCMR \\ \hline \hline
H1-1 & $8(3)\times10^{-5}$    & $2.6(4)\times10^{-3}$             & $2.76(1) \times 10^{-3}$   &  $<4\times10^{-5}$ \\ \hline
H1-2  & $5(2) \times 10^{-5}$    & $2.1(3) \times 10^{-3}$       & $3.2(2)\times 10^{-3}$ & $<6\times10^{-5}$  \\ \hline
\end{tabular}%
}
\caption{Typical errors measured in the two different quantum computers; H1-1 and H1-2. We use the following shorthand notation: single-qubit gate errors (SQ), two-qubit gate errors (TQ), state preparation and measurement errors (SPAM), mid-circuit measurement and reset crosstalk errors (MCMR). The SQ and TQ error rates are measured using randomized benchmarking experiments.}
\label{error_table}
\end{table}

The decoders used in these experiments were partially written in Rust and compiled to WebAssembly (Wasm). The stack (the entire system) has support for calling arbitrary Wasm functions via a foreign function interface (FFI) from QASM to Wasm. The choice of Wasm provides an efficient, safe (due to its sandbox), and portable classical language to have functions that are callable from quantum programs--where this language's code can be generated from many high-level language compilers. The resulting data returned from the classical functions are then used to dictate the control flow and operations executed in the quantum program.

Aside from requiring integers to be passed between Wasm and QASM and so long as the FFI function returns within a time set by the control system software, Wasm code is unconstrained in the internally-used classical features. For example, the decoder implemented in Rust exercises many high-level program constructs including arrays, tuples, higher-order functions, and match statements, that are all compiled to efficient Wasm. The support for these features means that various scalable algorithmic decoders can be ergonomically implemented in various high-level languages that compile to Wasm (such as Rust, C, and C++) and called from quantum programs.

\section{The five-qubit code}

\begin{figure*}[ht!]
\includegraphics[trim=15 0 0 0, clip,width=\textwidth]{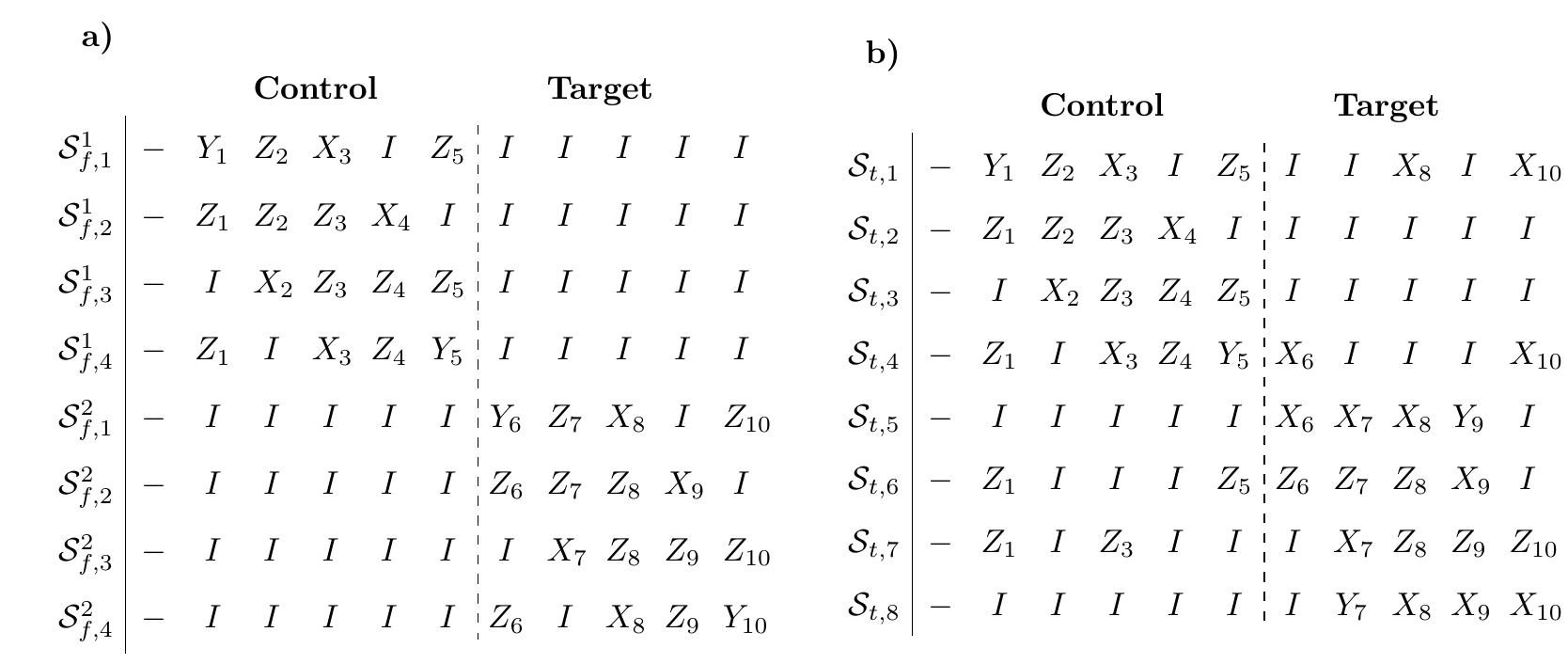}
\caption{The stabilizer generators of the joint $[[10,2,3]]$ code (\textbf{a}) before and (\textbf{b}) after the application of the first part of the pieceable fault-tolerant CNOT gate. Before the application of any physical-level entangling gates, the stabilizer generators are just two sets of the rotated five-qubit code stabilizers. The first subscript identifies which code the stabilizer generator belongs to, either $f$ for the five-qubit code or $t$ for the $[[10,2,3]]$ code, with the second subscript labeling the individual generator. In \textbf{a)}, the superscripts denote which logical qubit the generators belong to. After applying  physical-level entangling gates (the first two-thirds of the round-robin gate) the two sets of stabilizer generators become mixed to form new higher-weight stabilizer generators. The higher weight stabilizers are measured in the intermediate QEC cycle to maintain FT. The higher weight stabilizer generators can be derived from the original stabilizers by considering the action of the circuitry shown in Fig.~\ref{main_char}. For example, $\mathcal{S}_{t,6}=R_{2/3}U\mathcal{S}_{f,2}^2U^{\dagger}R_{2/3}^{\dagger}$, where $R_{2/3}$ denotes the first two-thirds (6 physical CNOTs) of the round-robin gate.}
\label{5_1_3_stabs_TQG}
\end{figure*}

The five-qubit code has a rich history in theoretical and experimental QEC studies as it was one of the first codes discovered and its minimal qubit resource requirement makes it suitable to small quantum computers. For example, the five-qubit code was used for the first experimental demonstration of QEC using an NMR platform~\cite{Knill2001} as well as the first FT gating operations~\cite{Zhang2012}. Recently the five-qubit code was used to demonstrate some basic necessities of QEC on superconductor~\cite{Gong2021} and diamond~\cite{Abobeih2021} systems. The code is defined by the stabilizer generators shown in Fig.~\ref{5_1_3_stabs_TQG}(a).

As mentioned previously, the five-qubit code does not admit a transversal CNOT gate, but an intriguing proposal known as pieceable FT~\cite{Yoder2016} does allow for a FT CNOT gate using only two logical qubits. The idea behind pieceable FT is to decompose an initially non-FT logical gate operation into pieces that are individually FT. With this decomposition, the pieces can be regarded as individual operations that can be bookended with QEC operations, thereby ensuring that faults do not have a chance to spread beyond the corrective power of the code.

In our study, the logical gate between two five-qubit code blocks using pieceable FT makes use of FT SPAM operations from \cite{Reichardt18} and an FT CNOT construction from \cite{Yoder2016}. We now describe the full FT procedure and will highlight deviations from this procedure later in the text.

The first step is a non-FT encoding circuit to initialize individual code blocks to logical $\ket{-}_L$ states. We then verify the state preparation of the logical states using a three-step procedure outlined in \cite{Reichardt18} that effectively measures different representations of the logical $\overline{X}$ operator, guaranteeing at most a weight-one error (\textit{i.e.}, one physical error) on the code block. In principle, this procedure can be run as a repeat-until-success circuit, as in Ref.~\cite{RyanAnderson2021}, ensuring in real time the state is FT prepared, but here we only ran this procedure once, and post-select on successful preparations. The probability of a successful FT initialization for the five-qubit code was found to be $0.909(1)\%$.

\begin{figure*}[ht!]
\includegraphics[trim=10 70 100 5, clip,width=\textwidth]{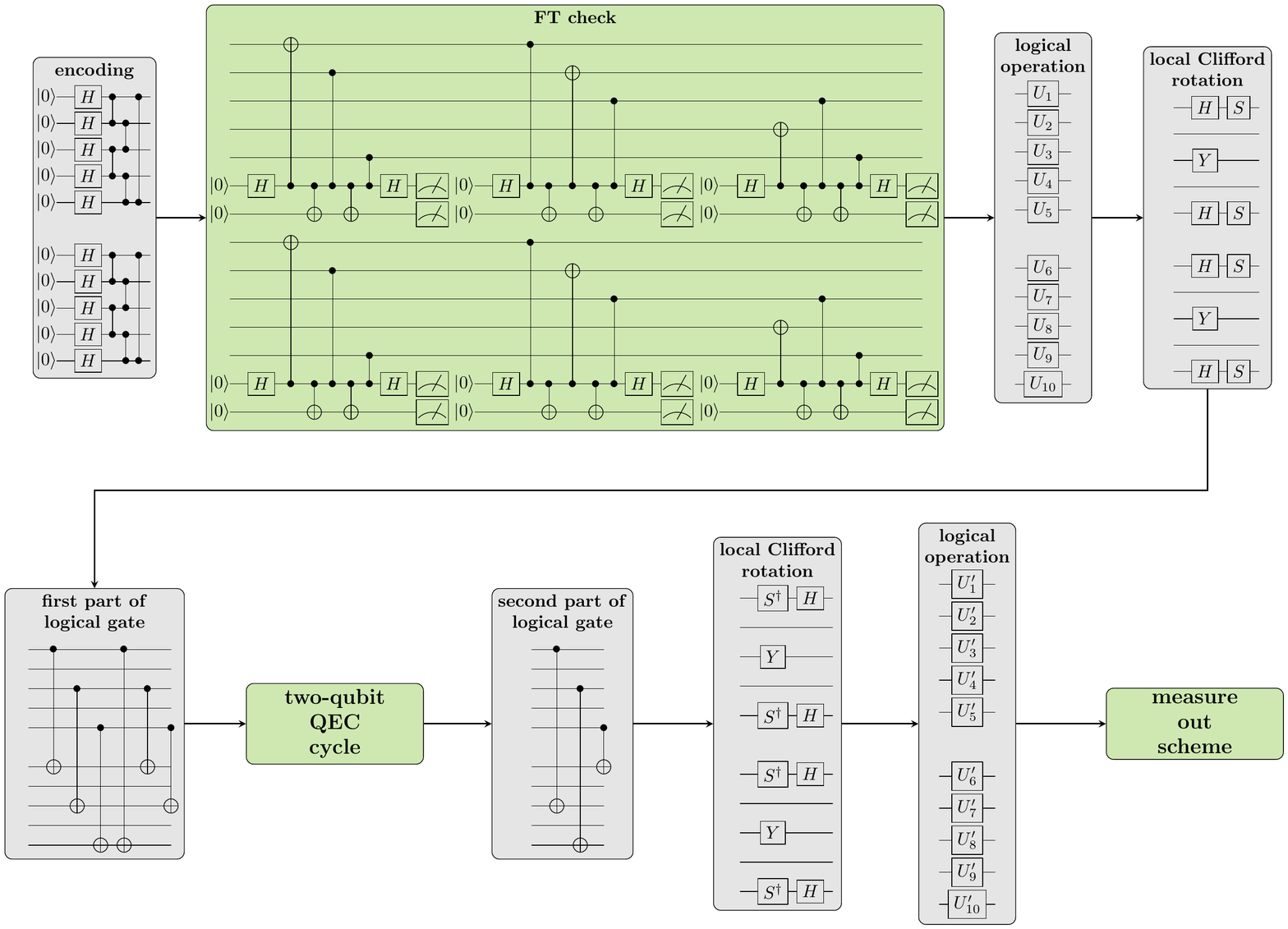}
\caption{The quantum circuit layout for the different CNOT constructions. The circuit elements are colored differently to indicate that some of the elements are constant in all of the constructions listed in Table \ref{FidelityTable_v1}, while some of them vary. All of the circuit elements colored gray are constant in each construction: Encoding, the first local Clifford rotation to prepare the input state to the gate, the first part of logical gate, the second part of logical gate, and the second local Clifford rotation which determines the measurement basis. The circuit elements that are colored green vary in our different constructions: fault-tolerant initialization, the two-qubit QEC cycle, and the measure-out scheme.}
\label{main_char}
\end{figure*}

Next, a logical operation rotates the logical basis state to the one we wish to prepare (\textit{e.g.}, $\ket{0}_L$ or $\ket{+}_L$). Then, because the construction of the pieceable FT gate in \cite{Yoder2016} relies on a different local Clifford equivalent version of the five-qubit code, we rotate to that version of the code by applying $U=S_1H_1Y_3S_5H_5$ to each code block (Fig. \ref{main_char}), where the phase gate $S=\textrm{diag}\{1,i\}$, $H$ is the standard Hadamard gate, and $Y$ is the Pauli-$Y$ operator. This new version of the code and its stabilizers allow for a straightforward construction of the ``round-robin" pieceable FT gate. This construction is adapted directly from \cite{Yoder2016} and differs mainly by implementing a logical CNOT instead of a logical CZ, which is achieved by applying nine physical CNOT gates instead of nine physical CZ gates. This nine gate circuit leaves the stabilizers of both code blocks unchanged and applies the appropriate action on the logical operators: $\overline{X}\overline{I}\rightarrow \overline{X}\overline{X}$, $\overline{I}\overline{X} \rightarrow \overline{I}\overline{X}$, $\overline{Z}\overline{I} \rightarrow \overline{Z}\overline{I}$, and $\overline{I}\overline{Z} \rightarrow \overline{Z}\overline{Z}$. All of this can be verified by conjugating the stabilizer and logical operators by the round-robin circuits.

The first piece of the round-robin gate proceeds by applying the first six CNOTs between qubits 1, 3, and 5 in each code block, (see Fig. \ref{main_char}). These CNOTs mix the stabilizers of the joint $[[10,2,3]]$ code. (Note, the $[[10,2,3]]$ stabilizers differ from those seen in~\cite{Yoder2016} due to applying physical CNOTs instead of physical CZs). We can then perform an intermediate two-qubit QEC cycle on this joint $[[10,2,3]]$ code, where a QEC cycle refers to a collection of multiple syndrome measurement rounds that are FT to measurement faults and where the syndromes from these rounds are decoded together to determine a correction~\cite{RyanAnderson2021}. 

The $[[10,2,3]]$ QEC cycle consists of measuring weight-4 and weight-6 stabilizers (see Figs. \ref{5_1_3_stabs_TQG} and \ref{qec_cycle_flag}) using flagged circuits to account for higher-weight hook errors. If a flag is triggered or a non-trivial stabilizer syndrome is measured, the QEC cycle is dynamically altered to measure the same stabilizers using the unflagged circuits (see Fig. \ref{qec_cycle_flag}). The combination of the flagged and unflagged syndrome measurements are then sent to individual look-up table decoders to determine what correction (if any) is needed~\cite{Reichardt18,RyanAnderson2021}. The corrections are determined and used to update a Pauli frame that is tracked in software, which is then used to correct the logical output at the end of the circuit execution. After the intermediate error correction step is finished, the last part of the logical CNOT gate is applied using three more physical CNOTs, again involving only qubits 1, 3, and 5 in each code block.

After the logical CNOT gate, we measure out the state in a FT manner~\cite{Reichardt18} by first rotating back to the canonical version of the code and then rotating to the appropriate measurement basis before applying a multi-step FT measurement procedure (see Fig. ~\ref{ft_meas} and Ref.~\cite{Reichardt18}). 

To characterize the impact of FT design and circuit depth, we ran five different experiments that used different combinations of circuit elements and we describe them here. 
\begin{itemize}
\item \textbf{SPAM1f} These SPAM fidelity experiments were done for the eigenstates of the logical $\overline{X}$ and $\overline{Z}$ operators using the non-FT encoding circuit, followed by a transversal single-qubit logical operation, and finished by a measurement (these steps are in the more general circuit in Fig.~\ref{main_char}, and the explicit circuit can be seen in Fig. \ref{nft_meas}). The $\overline{X}$ and $\overline{Z}$ bases resulted in respective fidelities of $0.978(2)$ and $0.976(2)$, respectively (see Table~\ref{FidelityTable_v1}).

\item \textbf{SPAM2f} These SPAM experiments  incorporated the FT verification step (see FT check in Fig.~\ref{main_char}) and a FT measure out scheme (Fig. ~\ref{ft_meas}) both described in \cite{Reichardt18}. Again, we find no significant difference between the logical $\overline{X}$ and $\overline{Z}$ SPAM fidelities, and find fidelities of $0.967(4)$ and $0.952(5)$, respectively (see Table~\ref{FidelityTable_v1}). 
\end{itemize}
We see that the non-FT SPAM (\textbf{SPAM1f}) procedure out performs the FT SPAM (\textbf{SPAM2f}) at current noise levels. We attribute this to the number of two-qubit gates involved in either procedure. The non-FT SPAM procedure has a total of 5 two-qubit gates per code block (see Fig. \ref{nft_meas}), whereas the full FT SPAM procedure has a vary number of two-qubit gates, due to the dynamical nature of the circuit, ranging from 30-40 two-qubit gates per code block in total. 

The next three experiments include a logical CNOT and different SPAM protocols. We apply the gate to an informationally-incomplete set of input states, and measure the fidelity of each output state. This precludes us from quoting a process fidelity estimate, but does allow for the derivation of rigorous bounds on the fidelity as discussed in Appendix~\ref{appendix_fidelity}. We use input states from three different bases: $XX$, $ZZ$ and $XZ$. The CNOT truth tables for these are: the $X$-basis
\begin{align}
\ket{++} &\rightarrow \ket{++} \nonumber \\
\ket{+-} &\rightarrow \ket{--} \nonumber \\
\ket{-+} &\rightarrow \ket{-+} \nonumber \\
\ket{--} &\rightarrow \ket{+-},
\label{eq.cxxrules}
\end{align}

\noindent the $Z$-basis
\begin{align}
\ket{00} &\rightarrow \ket{00} \nonumber \\
\ket{01} &\rightarrow \ket{01} \nonumber \\
\ket{10} &\rightarrow \ket{11} \nonumber \\
\ket{11} &\rightarrow \ket{10},
\label{eq.cxzrules}
\end{align}

\noindent and the mixed basis, which results in the creation of Bell states,
\begin{align}
\ket{+0} &\rightarrow \frac{1}{\sqrt{2}} \left(\ket{00} + \ket{11} \right) \nonumber \\
\ket{-0} &\rightarrow \frac{1}{\sqrt{2}} \left(\ket{00} - \ket{11} \right) \nonumber \\
\ket{+1} &\rightarrow \frac{1}{\sqrt{2}} \left(\ket{01} + \ket{01} \right) \nonumber \\
\ket{-1} &\rightarrow \frac{1}{\sqrt{2}} \left(\ket{01} - \ket{01} \right).
\label{eq.cxbellrules}
\end{align}

With these different input states, we made measurements using the following three instantiations of the CNOT gate: 
\begin{itemize}
\item \textbf{CNOT1f} We use non-FT SPAM circuits and the non-FT round-robin gate (without any quantum error correction) (see Fig. \ref{nft_TQG}).

\item \textbf{CNOT2f} We use FT SPAM circuits and the round-robin gate with a non-FT QEC cycle using only the unflagged circuits to measure one round of syndromes after the first six physical CNOTs.

\item \textbf{CNOT3f} We use FT SPAM and the round-robin gate with a fully FT QEC cycle embedded (see Figs. \ref{main_char} and \ref{qec_cycle_flag}).
\end{itemize}

\begin{figure}[ht]
\includegraphics[width=\columnwidth]{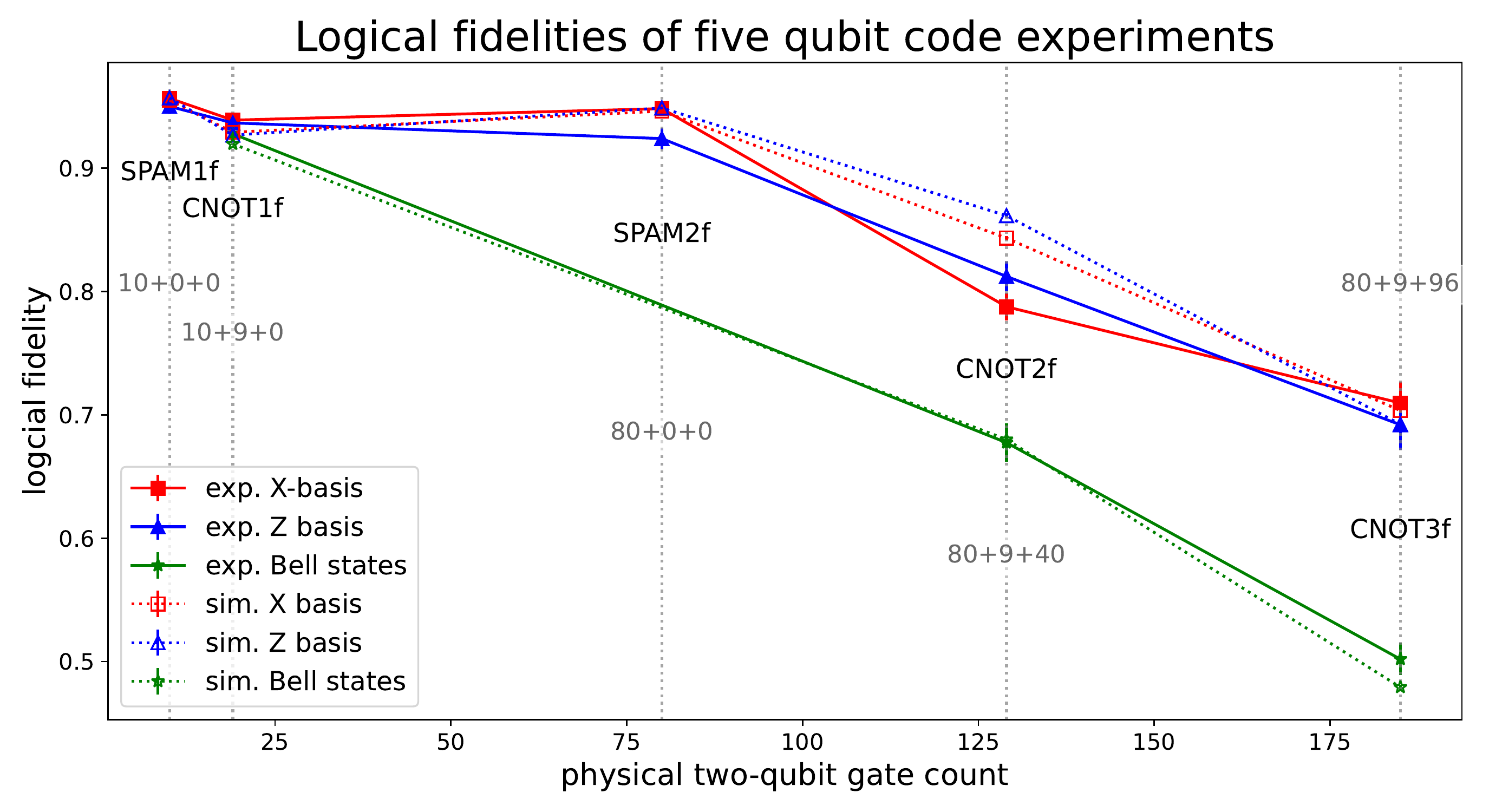}
\caption{Comparison of logical fidelity vs two-qubit gate count for five different five-qubit code experiments. Note that the ordering here is different than that in Table~\ref{FidelityTable_v1}, as denoted by the vertical line labels. Note that we also separate the experiments by the input state that was used, with red squares denoting the $\overline{X}$ basis states, blue triangles denoting $\overline{Z}$ basis states, and green stars denoting mixtures of $\overline{X}$ and $\overline{Z}$ basis states which ideally produce Bell pairs when acted upon by a CNOT gate.}
\label{fig_exp_comparison}
\end{figure}

The results from the experiments and subsequent analysis are in Table \ref{FidelityTable_v1} and Fig. \ref{fig_exp_comparison}. As Fig. \ref{fig_exp_comparison} shows, the extra circuitry designed to increase the degree of FT actually has a negative impact on the overall fidelity of the logical operation. Instead of the fidelity being correlated with the design principles to achieve FT, the fidelity is simply negatively correlated with the number of CNOT operations in the circuit. However, as we discuss further in Section~\ref{sec_simulations}, our simulations indicate this is likely not fundamental, but related to our specific noise environment, and the FT circuits may perform better in the presence of less dephasing. The highest fidelity state preparation, CNOT, measurement sequence is the non-FT sequence labeled \textbf{CNOT1f}, whose fidelity bounds are centered around $0.926$ as shown in Table~\ref{FidelityTable_v1}.

\begin{table*}[ht]
\begin{tabular}{|l|c|c|c|c|c|c|c|}
\hline
Circuit description       & Label & FT SPAM & [[10, 2, 3]] QEC & X-basis fidelity & Z-basis fidelity & Bell fidelity & Avg. fidelity bounds\\ \hline\hline
non-FT SPAM               & \textbf{SPAM1f} & no      & N/A             & $0.978(2)$ &  $0.976(2)$ & -           &    - \\ \hline
FT SPAM                   & \textbf{SPAM2f} & yes     & N/A             & $0.967(4)$ &  $0.952(5)$ &  -          &   -  \\ \hline
non-FT CNOT (no QEC)      & \textbf{CNOT1f} & no      & none            & $0.939(4)$ &  $0.937(4)$ & $0.928(4)$  & $[0.888, 0.948]$\\ \hline
non-FT CNOT (no flagging) & \textbf{CNOT2f} & yes     & 1 syn. extract. & $0.79(1)$  &  $0.81(1)$  &   $0.678(4)$ &   $[0.644, 0.848]$  \\ \hline
FT CNOT                   & \textbf{CNOT3f} & yes     & FT QEC cycle    & $0.71(2)$  &  $0.68(2)$  & $0.502(13)$   & $[0.467, 0.622]$ \\ \hline
\end{tabular}
\caption{The state fidelities achieved in the different five-qubit code experiments. The fidelity of the X-basis, Z-basis, and Bell states are found by preparing the input states listed in Eqs.~\ref{eq.cxxrules},\ref{eq.cxzrules},\ref{eq.cxbellrules} and measuring the appropriate corresponding logical operators. The bounds on the average fidelity with respect to CNOT listed in the rightmost column are then calculated according to the prescription outlined in Appendix \ref{appendix_fidelity}. The SPAM results in this table are calculated by preparing and measuring both qubits together, and we report the average of the two. This is in contrast to the logical SPAM fidelities plotted in Fig.~\ref{fig_exp_comparison}, which are the probabilities that both qubits were prepared correctly. Also note, the fidelities reported here are for the entire experiment; that is, the logical SPAM contribution has not been factored out of the fidelities reported for the logical CNOT experiments. 
The labels on the vertical lines correspond to the labels of the experiments in Fig.~\ref{fig_exp_comparison}.}
\label{FidelityTable_v1}
\end{table*}

\section{The color code}

\begin{figure*}[ht!]
\includegraphics[trim=15 275 60 0, clip,width=\textwidth]{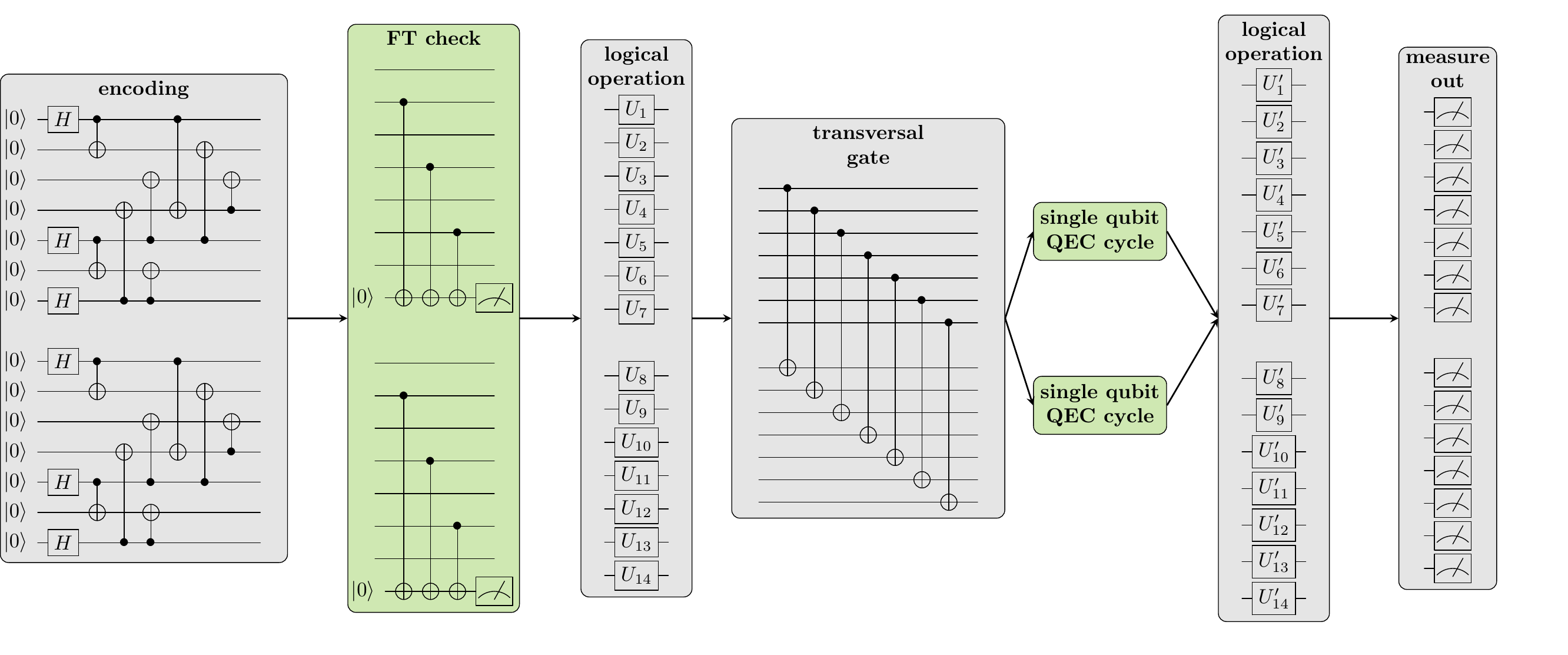}
\caption{The circuit used for testing the different instantiations of the color code's transversal CNOT gate. Using the same notation as with the five-qubit code circuit diagram, the grey boxes denote parts of the circuit that are used in all instantiations, and the green boxes denote additional circuitry used to increase the amount of FT.}
\label{713_trans_gate}
\end{figure*}

Similar to the five-qubit code, the color code has played an important role in the development of both theoretical and experimental QEC. Its early discovery~\cite{steane1996error} can be partially attributed to its simple construction using classical error correction codes, a construction whose generalizations are known as CSS (Calderbank-Shor-Steane) codes~\cite{Nielsen00}. It has since been realized that the code belongs to a family of 2D topological codes, known as color codes~\cite{bombin2006topological}. The code is defined by the stabilizer generators

\begin{flalign}
\nonumber
\mathcal{S}_{c,1}&=XXXXIII\\ \nonumber
\mathcal{S}_{c,2}&=IXXIXXI\\ \nonumber
\mathcal{S}_{c,3}&=IIXXIXX\\
\mathcal{S}_{c,4}&=ZZZZIII\\ \nonumber
\mathcal{S}_{c,5}&=IZZIZZI\\ \nonumber
\mathcal{S}_{c,6}&=IIZZIZZ. \nonumber
\label{CC_stabs}
\end{flalign}
where we use the subscript $c$ to distinguish the generators of the color code from the five-qubit, and the numerical subscript to label the individual generators. 

The color code has several nice properties, the most relevant being the admission of a transversal CNOT gate. Given the natural FT of the color code transversal CNOT gate, we cannot make a direct comparison between it and the different instantiations of the five-qubit code gate. However, it is possible to test similar effects related to different instantiations of the more general circuit that includes SPAM and additional QEC cycles.

The color code experiments were implemented in a similar fashion as described in our previous work, Ref.~\cite{RyanAnderson2021}. In the present work, however, only a single round of initialization was used to most closely match the design of the five-qubit code experiments. Experiments were then post-selected based on the results of the validation step. The probability of a successful FT initialization verification for the code code was found to be $0.9834(3)$. Note, we do not expect significant change the logical SPAM fidelity if we utilize a repeat-until-success method with no post-selection as done in Ref.~\cite{RyanAnderson2021} since in that method qubits are reset and the dephasing noise in the system imparts a logical Z error whch does not affect the $Z$-basis.

\begin{figure}[ht]
\includegraphics[width=\columnwidth]{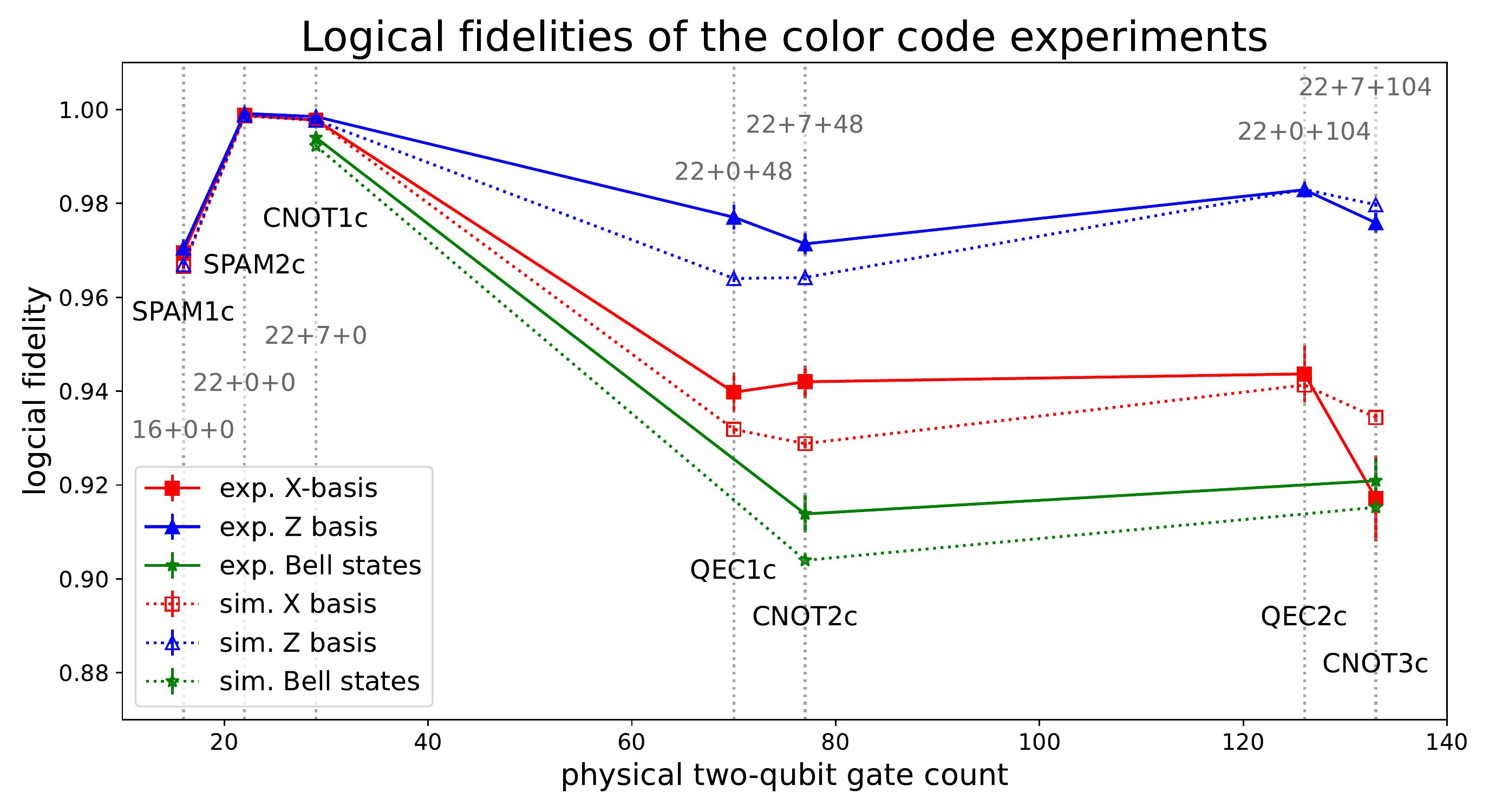}
\caption{Experimental and simulated logical fidelity vs the total two-qubit gate count in seven different color code experiments. Note that the ordering here is different than that in Table~\ref{FidelityTable_steane}, as denoted by the vertical line labels. The red, blue and green markers denote experiments that ideally produce output states in the $\overline{X}$, $\overline{Z}$, and Bell bases respectively. The markers connected with solid lines represent experimental data, and the markers connected with dashed lines represent simulated data.}
\label{fig_exp_comparison_steane}
\end{figure}

We ran seven different experiments to characterize the FT operation of the color code and we describe them here.
\begin{itemize}
\item \textbf{SPAM1c:} These experiments measured SPAM fidelities for the eigenstates of the logical $\overline{X}$ and $\overline{Z}$ operators using the non-FT encoding circuit, followed by a transversal single logical operation, and finished by a measurement (these steps are in the more general circuit in Fig.~\ref{713_trans_gate}). The $\overline{X}$ and $\overline{Z}$ bases resulted in respective fidelities of $0.9847(8)$ and $0.9852(9)$ (Table~\ref{appendix_fidelity}).

\item \textbf{SPAM2c:} These experiments measured SPAM fidelities for the eigenstates of the logical $\overline{X}$ and $\overline{Z}$ operators and included the FT check circuit shown in Fig.~\ref{713_trans_gate}, followed by a transversal single logical operation, and finished by a measurement. The $\overline{X}$ and $\overline{Z}$ bases resulted in respective fidelities of $0.99939(15)$ and $0.99959(13)$ (Table~\ref{appendix_fidelity}).

\item \textbf{QEC1c:} These experiments used the FT SPAM circuits and included a single non-FT QEC cycle on each qubit.

\item \textbf{QEC2c:} These experiments used the FT SPAM circuits and included a FT QEC cycle on each qubit.

\item \textbf{CNOT1c:} These experiments used FT SPAM circuits, included the transversal CNOT gate, but no QEC circuitry.

\item \textbf{CNOT2c:} These experiments used FT SPAM circuits, included the transversal CNOT gate, and a single non-FT QEC cycle on each qubit.

\item \textbf{CNOT3c:} These experiments used FT SPAM circuits, included the transversal CNOT gate, and a single FT QEC cycle on each qubit.
\end{itemize}
In contrast to our experiments with the five-qubit code, we find the fidelity of the different color code circuits is not simply negatively correlated with the number of physical two-qubit gate operations (Table~\ref{FidelityTable_steane} and Fig.~\ref{fig_exp_comparison_steane}). In particular, the SPAM circuits benefit largely from the addition of FT circuitry, seeing the error rates drop by roughly an order of magnitude. Technically speaking, this is the only additional circuitry that is required to make the CNOT circuit end-to-end FT, since the logical CNOT is transversal and naturally FT. For comparisons sake, as mentioned before we add additional QEC cycles and indeed find that the overall error rate of the circuit increases.
   
\begin{table*}[ht]
\begin{tabular}{|l|c|c|c|c|c|c|c|}
\hline
Circuit description  & Label  & FT SPAM & QEC rounds      & X-basis fidelity & Z-basis fidelity & Bell fidelity & Avg. fidelity bounds \\ \hline\hline
non-FT SPAM          & \textbf{SPAM1c} & no      & N/A             & $0.9847(8)$      &  $0.9852(9)$     & -             &     -    \\ \hline
FT SPAM              & \textbf{SPAM2c} & yes     & N/A             & $0.99939(15)$    &  $0.99959(13)$   & -             &     -    \\ \hline
non-FT QEC           & \textbf{QEC1c}  & yes     & 1 syn. extract. & $0.970(2)$       &  $0.988(1)$      & -             &     -                  \\ \hline
FT QEC               & \textbf{QEC2c}  & yes     & 1 FT QEC cycle  & $0.9711(31)$     &  $0.9914(8)$     & -             &     -                  \\ \hline
FT CNOT              & \textbf{CNOT1c} & yes     & none            & $0.9978(5)$      &  $0.9985(4)$     & $0.9940(7)$   &    $[0.9957, 0.9963]$    \\ \hline
FT CNOT + non-FT QEC & \textbf{CNOT2c} & yes     & 1 syn. extract. & $0.942(3)$       &  $0.971(2)$      & $0.914(4)$    &  $[0.9216, 0.9373]$    \\ \hline
FT CNOT + FT QEC     & \textbf{CNOT3c} & yes     & 1 FT QEC cycle  & $0.917(9)$       &  $0.976(2)$      & $0.921(5)$    &  $[0.8983, 0.9436]$    \\ \hline
\end{tabular}
\caption{The state fidelities achieved in the different code experiments. The results given in this table are for the entire experiment and do not have the logical SPAM fidelity factored out. The SPAM and one QEC cycles results in this table are calculated by preparing and measuring both qubits together, and we report the average of the two. The fidelity reported for one QEC cycle corresponds to applying one QEC cycle to both of the logical qubits in parallel. The bounds on the average fidelity with respect to CNOT in the rightmost column are derived using the method outlined in Appendix \ref{appendix_fidelity}.}
\label{FidelityTable_steane}
\end{table*}

It is worth noting the difference between these results and those of Ref.~\cite{RyanAnderson2021}. In our previous work, the logical SPAM error was $1.7(4)\times10^{-3}$, which was not significantly better than the physical SPAM error rate. The experimental upgrades mentioned previously (along with several other technical improvements to lower most of the physical operation error rates) have improved the average logical SPAM error to $0.5(2)\times10^{-3}$, which is significantly better than the current average physical SPAM error of $2.76(1)\times10^{-3}$.

Upon including the logical CNOT operation, the best fidelity we achieve is for the \textbf{CNOT1c} sequence. This sequence does not contain any QEC cycles but is still FT as a consequence of the transversal circuit structure between the FT initialization and measurement circuitry. We find the highest fidelities when the ideal output state is a product state and lower fidelities when the ideal output is an entangled state. As shown in Table~\ref{FidelityTable_steane}, the experimental sequence of FT initialization, transversal CNOT, and a FT measure out scheme, generated average fidelity bounds $[0.9957,0.9963]$ or an error rate of $\sim4\times10^{-3}$. It is compelling to compare this with the average fidelity of these same operations at the physical level, which were measured to be $[0.9850,0.9903]$, or an error rate of at least $\sim1.0\times10^{-2}$.

As discussed in Appendix~\ref{appendix_fidelity}, we can estimate the average fidelity bounds for the physical gate in a way that corrects for SPAM errors. The procedure we use assumes the SPAM error is dominated by the measurement error, which is well justified at the physical level~\cite{Olmschenk07}. At the physical level, we find SPAM-corrected bounds of $[0.9947,0.9957]$, which are still lower than the logical bounds that include SPAM error. The assumption of measurement noise being dominant at the logical level is less well justified, which is why we focus on the average fidelity bounds for the logical CNOT without correcting for SPAM. To
estimate the average fidelity of the logical gate in a
SPAM-independent way would require randomized benchmarking
or gate set tomography~\cite{Kohout2013}.

These results do not suggest our system is capable of executing arbitrary algorithms at the logical level with higher fidelity than algorithms executed at the physical level, or that we have achieved the so-called break-even point. Nevertheless, these measurements conclusively show that the sequence of state preparation-CNOT-measure operations is done with higher fidelity at the logical level than at the physical level, marking an important milestone in the march toward the break-even point. However, we note that the inclusion of QEC cycles along with more careful measurements will be crucial components in a ``fair'' comparison between the performance of physical and logical qubits.

\section{Simulations}
\label{sec_simulations}
To further understand the implications of our measurements, we performed detailed simulations of these systems. Similar to Ref.~\cite{RyanAnderson2021}, the simulations are parameterized by several physical-level error sources, most of which are independently characterized. Gate errors are estimated with single- and two-qubit randomized benchmarking, SPAM error is measured by standard single-qubit state preparation and measurement experiments, and measurement crosstalk is characterized by bright-state depumping rate measurements~\cite{qtuumspec}. The transport times used to generate dephasing errors were gathered by the stack's compiler acting on the quantum circuits to produce the real transport, cooling, and gating times.

To accurately account for dephasing errors in the emulator, qubits would need different frequency errors in different portions of the circuit since they travel to different trap locations through the course of the computation, but our emulator currently does not capture this. Additional complications include imperfections in the H1-1 phase tracking routine and time-dependent magnetic field noise. Since these nuances are not fully characterized, our emulator uses a dephasing rate variable that is constant throughout the circuit, and acts as a tunable fitting parameter. We use a coherent dephasing model (instead of incoherent) ~\cite{RyanAnderson2021} as coherent noise is more damaging, providing more conservative performance estimates. This model was then compared to measured data to find effective dephasing rates that best describe the experiments, (see Appendix~\ref{appendix_dephasing}). The two quantum computers have slightly different magnetic field environments, with H1-2 measuring qubit frequency fluctuations on the order of $0.4$ Hz, corresponding to $120~\mu$G ~\cite{Olmschenk07}, and H1-1 measuring inhomogeneities on the order of $2.5$ Hz, but the spatial phase tracking compensation results in more comparable effective dephasing rates. Using the method described in Appendix~\ref{appendix_dephasing}, we find effective dephasing rates of 0.24 Hz and 0.17 Hz in H1-2 and H1-1, respectively.

\begin{figure}[ht]
\includegraphics[width=\columnwidth]{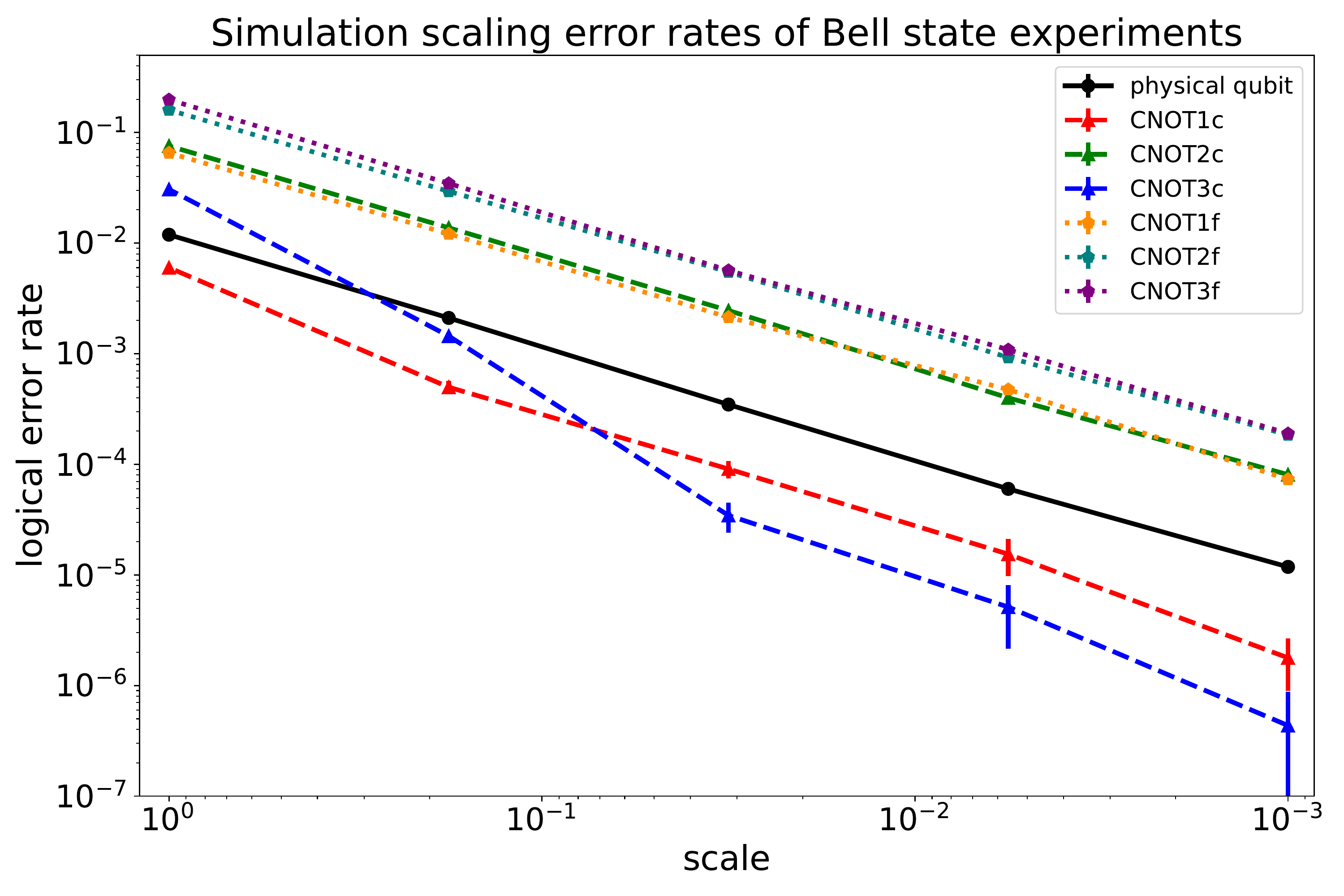}
\caption{Scaling physical error rates to determine the impact on logical error rate of the Bell state experiments for the five-qubit and color codes. The unscaled (scale = 1) physical error rates are those currently measured for H1-1 \cite{qtuumspec} with the exception that the dephasing rate starts at 0.08 Hz and leakage events are replaced by completely depolarizing noise to approximate the effect of leakage repumping~\cite{hayes2020eliminating}. The dephasing rate of 0.08 Hz was used as it was determined in Ref.~\cite{RyanAnderson2021} to be approximately where the logical error rate of a QEC cycle would be equal to the largest physical error rate which stems from the two-qubit gate.}
\label{fig_scaling_bell}
\end{figure}

With the informed emulator in hand, we performed numerical experiments to probe how the two codes behave in lower-error regimes where one might expect the logical encoding to outperform the physical operations. This calculation is done staring with a simplified set of error parameters and error model as described in caption of Fig.~\ref{fig_scaling_bell} and then scaling all error rates linearly together. As can be seen in Fig.~\ref{fig_scaling_bell}, the five-qubit code error rates are larger than the physical error rates for scaling factors as low as $10^{-3}$, (corresponding to a physical two-qubit gate error of $\sim2\times10^{-6}$). In fact, we find that the logical error scales \textit{nearly} linearly within the range of physical probed, indicating the code is being operated far from its pseudo-threshold value.

Our results do not rule out using the five-qubit code for quantum computing, as we are only investigating gating operations here and other studies have shown the code to perform well as a quantum memory~\cite{Xu2018}. Additionally, the implementation of the logical two-qubit gate protocol maybe be further optimized as we are only utilizing two ancillas and there maybe more optimization to be done at the transport and/or circuit level. Further, the round-robin gate scheme employed here is not the only FT method of implementing entangling gates between five-qubit code logical qubits. Gottesman~\cite{Gottesman1998} gave a construction using a 3-qubit transversal gate on the five-qubit code which can form the basis for a CNOT gate with the addition of measurements and single qubit operations. We leave the study of this logic to future work.

The simulation data in Fig.~\ref{fig_scaling_bell} indicates that for low enough physical error rates, the color code CNOT with an added FT QEC cycle should eventually outperform the standalone gate, but requires somewhat lower error rates than we currently achieve. In contrast, adding a \textit{non}-FT QEC to the end of the gate operation causes the simulated logical gate to always perform worse than the physical operation in the error regimes we probed.

\section{Discussion and Conclusion}
In this work, we have presented the first experimental comparison study of different QEC codes in similar environments. Using real-time classical co-processing, we implemented FT protocols to entangle logical qubits with high fidelity, demonstrating the feasibility of using QEC for error suppression in quantum information processing once the physical error rates are sufficient. Our study is not exhaustive, but it is suggestive that the relatively economical FT circuitry of the color code will provide a better platform for computation than the qubit efficient five-qubit code. For example, the two codes perform similarly in the \textbf{SPAM1f} and \textbf{SPAM1c} experiments which are only encoding circuits followed by measurements. However, making these SPAM circuits FT requires a much larger circuit in the case of the five-qubit code whereas the color code has a very modest overhead. As a result, the five-qubit code FT SPAM circuit performs worse than the \textit{non}-FT encoding, whereas the color code FT SPAM circuit has an order of magnitude smaller error than the non-FT encoding circuit. 

We caution that this conclusion is limited to the specific protocols studied here and we leave the study of other FT protocols to future work. Likewise, we also caution that the comparison of the protocols themselves is limited to these codes~\textemdash~ in particular, the concept of pieceable fault tolerance is a very general concept, having many applications in other codes and is likely to be a valuable tool in future studies.

Intriguingly, during the course of this trade study, we found indications of a logical two-qubit gate break-even milestone. The color code transversal CNOT experiment \textbf{CNOT1c} found higher average fidelity bounds than the analogous physical level experiment. The increased fidelity of the logical circuit is mostly due to the SPAM operations being higher fidelity at the logical level as compared to the physical level, but even after correcting for SPAM we find the logical CNOT operation is higher fidelity than the physical operation. To our knowledge, this is the first time this milestone has been achieved for a logical two-qubit-gate. An important next step would be to measure the logical gate fidelity in a SPAM independent way. We additionally note that the error measured when QEC cycles are added is significantly higher (approximately $1\%$) than the physical two-qubit gate, highlighting another key milestone that must be achieved before logical algorithms can be expected to outperform unencoded computations. With the general goal of logical quantum computing being to implement quantum algorithms with higher logical fidelity, systems will need to maintain good performance in both logical gates \textit{and} logical memory as they are scaled up. Thus, the results presented here can be seen as one in a series of milestones towards the overall goal of beneficial logical quantum computation.

This work is just the beginning of experimental trade studies into what QEC code fits best with our QCCD architecture. One feature of the QCCD architecture is that qubits are mobile, meaning circuit geometries do not have to match the hardware geometry. Most proposals for QCCD computers envision using 2D surface traps compatible with scalable micro-fabrication processes, which may be designed to match 2D code geometries to optimize clock speed, but it is possible that higher dimensional codes offer benefits that eclipse those of 2D codes. For example, recent progress has been made in showing that low-density parity check codes can achieve high encoding rates and may offer lower physical resource overheads~\cite{Panteleev2021,Lin2022,Cohen2022,Krishna2021}.

It is currently difficult to predict which codes and implementations of those codes may perform the best in general scenarios, and when considering anything but the simplest error models, one is usually forced to resort to numerical studies. Additionally, the exploration space is vast. There are many families of codes to consider and different schemes for gating, encoding, decoding and syndrome extraction. On top of that, one can consider designing different device geometries that are best suited to each option. It is likely not feasible to experimentally explore all instantiations one at a time, and we hope studies like ours will help researchers understand what the most important properties are, and begin to hone in on truly practical quantum error correction.

\section{Acknowledgements}
We acknowledge the countless contributions and work of the entire team at Quantinuum. We'd also like to thank Ben Criger and Isaac Kim for helpful comments on the manuscript. These experiments were done using the Quantinuum system models H1-1 and H1-2, both powered by Honeywell ion traps.

\bibliography{inversion}
\appendix

\section{Process fidelity bounds from semidefinite programming}
\label{appendix_fidelity}
In this appendix, we review how to bound the
quantum process fidelity with respect to an
ideal target unitary, from a set of state fidelities
obtained from applying the process to each state. 
This procedure was originally described in Ref.~\cite{Audenart2002}.
For cleaner notation, we drop the overbars and subscripts denoting logical operators and states.
We will make use of the Choi matrix formalism.
Let $\mathcal{H}=\mathbb{C}^d$ be the Hilbert space for a $d$-dimensional quantum system,
and let $\mathcal{B}(\mathcal{H})$ denote the space of linear operators on $\mathcal{H}$.
Let $\mathcal{E}$ be a quantum process, that is, a completely-positive
trace-preserving (CPTP) map on $\mathcal{B}(\mathcal{H})$.
Then the Choi matrix  of $\mathcal{E}$ is defined as
\begin{equation}
    \chi_{\mathcal{E}}=(\mathcal{I}\otimes\mathcal{E})(\ket{\phi}\bra{\phi}),
\end{equation}
where $\mathcal{I}$ is the identity process, and $\ket{\phi}$ is
a maximally entangled state in $\mathcal{H}\otimes\mathcal{H}$ given by
\begin{equation}
    \ket{\phi}=\frac{1}{\sqrt{d}}\sum_{i=0}^{d-1}\ket{i}\otimes\ket{i}.
\end{equation}
The output of $\mathcal{E}$ on any state $\rho$ is given by
\begin{equation}
    \mathcal{E}(\rho)=d\Tr_1\big(\chi_{\mathcal{E}}(\rho^{\intercal}\otimes I)\big),
\end{equation}
where $\intercal$ denotes transposition, and the partial trace
is over the first subsystem.
Indeed, this can be shown as follows:
\begin{align}
    d\Tr_1\big(\chi_{\mathcal{E}}(\rho^{\intercal}\otimes I)\big) &=
    \sum_{i,j}\Tr_1\big((\ket{i}\bra{j}\otimes\mathcal{E}(\ket{i}\bra{j}))(\rho^{\intercal}\otimes I)\big)\notag\\
    &=\sum_{i,j}\bra{j}\rho^{\intercal}\ket{i}\mathcal{E}(\ket{i}\bra{j})\notag\\
    &= \sum_{i,j}\bra{i}\rho\ket{j}\mathcal{E}(\ket{i}\bra{j})\notag\\
    &= \mathcal{E}(\rho).
\end{align}
The map $\mathcal{E}$ is completely-positive if and only if $\chi_{\mathcal{E}}$ is positive semidefinite~\cite{Choi1975}.
Also, $\mathcal{E}$ is trace-preserving if and only if $d\Tr_{2}(\chi_{\mathcal{E}})=I$. 

Now we introduce the $\textit{process fidelity}$,
which is sometimes also called the entanglement fidelity~\cite{Schumacher1996}.
If $U$ is a unitary, then the process fidelity of $\mathcal{E}$ with
respect to $U$ is
\begin{align}
    F(\mathcal{E},U)&=\bra{\phi}(I\otimes U^{\dag})\big(\mathcal{I}\otimes\mathcal{E}\big)(\ket{\phi}\bra{\phi})(I\otimes U)\ket{\phi}\notag\\
    &= \Tr(\chi_{\mathcal{E}}\chi_{U}).
\end{align}
A related quantity is the average fidelity,
$F_{\mathrm{avg}}(\mathcal{E}, U)$, defined as
\begin{equation}\label{eq: average fidelity}
    F_{\mathrm{avg}}=\int d\psi\bra{\psi}U^{\dag}\mathcal{E}(\ket{\psi}\bra{\psi})U\ket{\psi},
\end{equation}
which is computed from the process fidelity by the formula~\cite{Nielsen2002}
\begin{equation}
    F_{\mathrm{avg}}=\frac{d\times F+1}{d+1}.
\end{equation}
The reported fidelity bounds in
tables~\ref{FidelityTable_v1}, ~\ref{FidelityTable_steane}, and~\ref{PhysicalFidelityTable}
are given in terms of the average fidelity.

Suppose that for some set of states $\{\ket{\psi_a}\}_a$
the output of the process on each state satisfies
\begin{equation}
    \bra{\psi_a}U^{\dag}\mathcal{E}(\ket{\psi_a}\bra{\psi_a})U\ket{\psi_a}= f_a.
\end{equation}
In terms of the Choi matrix , this can be rewritten as
\begin{equation}\label{eq: SDP constraints}
    \Tr(\chi_{\mathcal{E}}(\ket{\psi_a}\bra{\psi_a}^{\intercal}\otimes U\ket{\psi_a}\bra{\psi_a}U^{\dag}))= \frac{1}{d}f_a.
\end{equation}
The minimum process fidelity among all processes consistent with~\eqref{eq: SDP constraints} is given by the solution to the optimization problem
\begin{align}\label{eq: SDP}
    \mathrm{Minimize}: &\Tr(\chi_{\mathcal{E}}\chi_{U})\notag\\
    \mathrm{Subject\,to}:&\Tr(\chi_{\mathcal{E}}A_a)=\frac{1}{d}f_a\notag\\
    &\Tr(\chi_{\mathcal{E}}B_{ij})=\frac{1}{d}\delta_{ij}\notag\\
    &\chi_{\mathcal{E}}\ge 0.
\end{align}
Here $A_a=\ket{\psi_a}\bra{\psi_a}^{\intercal}\otimes U\ket{\psi_a}\bra{\psi_a}U^{\dag}$,
and $B_{ij}=\ket{i}\bra{j}\otimes{I}$,
which enforces the partial trace constraint since $\Tr_2(\chi_{\mathcal{E}})_{ij}=\Tr(\chi_{\mathcal{E}}B_{ij})$.
The optimization problem~\eqref{eq: SDP} is in the standard form of a
semidefinite program~\cite{Vandenberghe1996}.
The maximum process fidelity of
all processes satisfying~\eqref{eq: SDP constraints}
is given by the solution to an analogous SDP.

In our experiment, the input states $\{\ket{\psi_a}\}_a$ are the $X$-basis states,
$Z$-basis states, and Bell state preparation states
given by the left hand sides of Eqs.~\eqref{eq.cxxrules}-\eqref{eq.cxbellrules}.
For the $X$ and $Z$ basis states,
output state fidelities $f_a$ are equal
to the probabilities of measuring the correct
output state.
To obtain the Bell state fidelities,
the two logical qubits must be measured in the $XX$, $YY$, and $ZZ$ bases.
The state fidelity with respect to the Bell
state $\ket{\psi_{\mathrm{Bell}}}=\frac{1}{\sqrt{2}}(\ket{00}+\ket{11})$ is given by
\begin{equation}
    \bra{\psi_{\mathrm{Bell}}}\rho\ket{\psi_{\mathrm{Bell}}}=\frac{1}{4}\big(1+\langle XX\rangle-\langle YY\rangle + \langle ZZ\rangle\big),
\end{equation}
where the bracket denotes expectation value.
The fidelity with respect to the other Bell states in Eq.~\eqref{eq.cxbellrules} are given by similar expressions.

From the experimental estimates $\{\hat f_a\}_a$ of the output state fidelities,
and their associated statistical errors $\{\epsilon_a\}_a$,
we replace the first constraint in~\eqref{eq: SDP} with $\Tr(\chi_{\mathcal{E}}A_a)\ge\frac{1}{d}(\hat f_a-\epsilon_a)$,
and with $\Tr(\chi_{\mathcal{E}}A_a)\le\frac{1}{d}(\hat f_a+\epsilon_a)$
for the analogous maximization problem. 
We solve the SDP given by~\eqref{eq: SDP}
using the CVXPY software package~\cite{Diamond2016, Agrawal2018}.

The procedure as described so far assumes
perfect state preparation and measurement (SPAM).
For the fidelity bounds on the physical gates,
we correct for SPAM errors as follows.
At the physical level, measurement errors are much larger than state 
initialization and single-qubit gate errors,
so we can reasonably assume perfect state prep,
and use the results of the physical SPAM experiments
to estimate the POVM elements corresponding to each measurement outcome.
As a specific example, if $\{E_0, E_1\}$ is the POVM
corresponding to a single-qubit $Z$-basis measurement,
then
\begin{equation}
    E_0 = \begin{pmatrix}
        p(0|0) & 0\\
        0 & p(0|1)
        \end{pmatrix},\,
    E_1 = \begin{pmatrix}
        p(1|0) & 0\\
        0 & p(1|1)
        \end{pmatrix},
\end{equation}
where $p(i|j)$ is the probability of outcome $i$ given
state preparation $j$.
Note that $E_0+E_1=I$, as required by a POVM.
Similarly, for the $X$-basis measurement,
\begin{equation}
    E_+ = \begin{pmatrix}
        p(+|+) & 0\\
        0 & p(+|-)
        \end{pmatrix},\,
    E_- = \begin{pmatrix}
        p(-|+) & 0\\
        0 & p(-|-)
        \end{pmatrix}.
\end{equation}
The results of the physical SPAM experiments
are summarized in Table~\ref{PhysicalFidelityTable},
where the $X$-basis SPAM error is the mean of $p(+|-)$ and $p(-|+)$,
and similarly the $Z$-basis SPAM error is the
mean of $p(0|1)$ and $p(1|0)$.
From the single-qubit POVM elements,
the two-qubit POVM elements are computed as $E_{ij}=E_i\otimes E_j$ for $i,j\in\{0,1,+,-\}$.
Using estimates of the noisy measurement operators,
the operator $A_a$ in the SDP of Eq.~\eqref{eq: SDP} is given
by $A_a=\ket{\psi_a}\bra{\psi_a}^{\intercal}\otimes U E_a U^{\dag}$,
where $E_a$ is the noisy two-qubit POVM element that in the
limit of zero noise equals $\ket{\psi_a}\bra{\psi_a}$.

\begin{table*}[ht]
\begin{tabular}{|c|c|c|c|c|c|c|}
\hline
X-basis SPAM & Z-basis SPAM & X-basis fidelity & Z-basis fidelity & Bell fidelity & Avg. fidelity bounds & Avg. fidelity bounds (SPAM corrected)\\ \hline \hline
0.9970(2) & 0.9985(1) & 0.9913(8) & 0.9921(4) & 0.9870(5) & [0.9850, 0.9903] & [0.9947, 0.9957] \\ \hline
\end{tabular}
\caption{SPAM fidelities, output state fidelities, and average fidelity bounds for the physical CNOT gate.
Each experiment was run on the Quantinuum H1-1 system, in parallel across all
5 active gate zones, and the numbers in the table are averaged
over all zones.
The average fidelity bounds uncorrected for SPAM errors are obtained 
by solving the SDP of Eq.~\eqref{eq: SDP} with ideal state preparation and measurements.
The SPAM corrected average fidelity bounds are 
obtained using noisy measurement operators inferred
from the SPAM experiments, as described in Appendix~\ref{appendix_fidelity}.
}
\label{PhysicalFidelityTable}
\end{table*}

\section{Additional five-qubit code details}

Fig.~\ref{5_1_3_stabs_TQG} details the five-qubit code stabilizer generators, as well as how they are transformed by the pieceable round-robin gating operation. Here, we provide additional details about the code structure.

In order for quantum error correction to be useful, we not only have to be able to detect errors via syndrome extraction and decoding, but then we have to be able to apply recovery operations to the data qubits. Recovery operations can be constructed using a product of elements from three separate groups: the logical operators $\mathcal{L}$, the stabilizers $\mathcal{S}$, and the destabilizers $\mathcal{D}$.

The logical operators, which also serve as generators of the logical group, are

\begin{align}
\overline X_{f}&=-YIXIY\\
\overline Y_{f}&=-ZIYIZ\\
\overline Z_{f}&=-XIZIX.
\end{align}

\noindent The stabilizers generators are seen in Fig.~\ref{5_1_3_stabs_TQG} but for completeness if given as

\begin{align}
\mathcal{S}_{f,1}&=XZZXI\\
\mathcal{S}_{f,2}&=IXZZX\\
\mathcal{S}_{f,3}&=XIXZZ\\
\mathcal{S}_{f,4}&=ZXIXZ.
\end{align}

\noindent The destabilizer generators for the five-qubit code is given by
\begin{align}
\mathcal{D}_{f,1}&=IXIII\\
\mathcal{D}_{f,2}&=YXIIY\\
\mathcal{D}_{f,3}&=XIIIX\\
\mathcal{D}_{f,4}&=IXIZI.
\end{align}

\noindent The a Pauli recovery operations $P$ can always be decomposed as some element $P \in \mathcal{L}\times \mathcal{S} \times \mathcal{D}$. Note, in the case of the FT round-robin construction, just like the stabilizers are transformed into different operators due to the application of two-qubit gates, the destabilizers also be transformed. The transformations of both destabilizers and the Pauli frame must be track to determine the appropriate recovery operation and to derive changes in syndromes so that previous syndrome measurements can be properly modded out in subsequent measurements of syndromes and logical representatives.

\section{Dephasing simulations}
\label{appendix_dephasing}

\begin{figure}[ht]
\includegraphics[width=0.85\columnwidth]{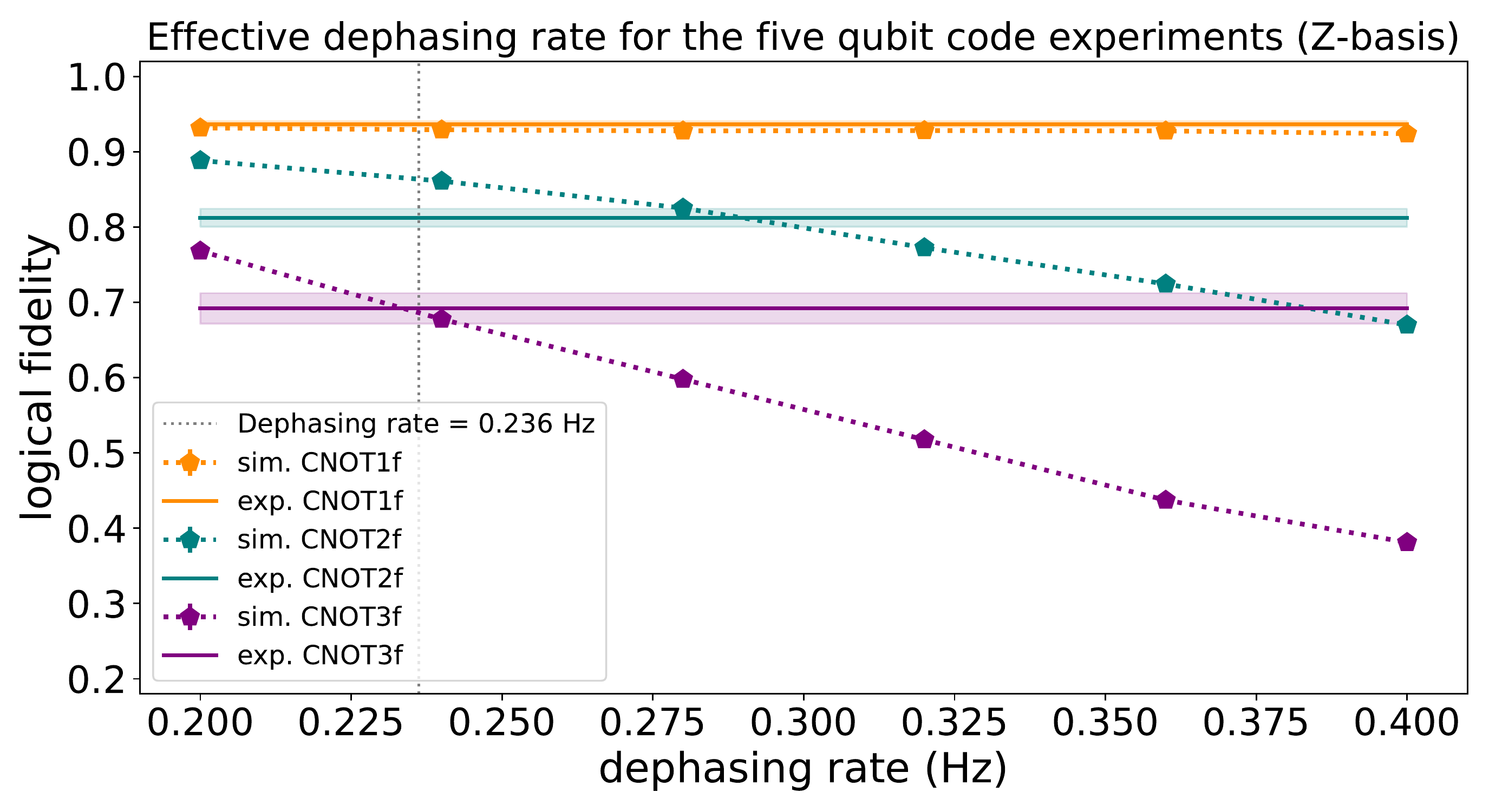}
\caption{Determination of dephasing rate parameter for simulations of the five-qubit code using logical CNOT experiments for the H1-2 device. Shown here are the experimental measurements $m_i$ and the simulated results $s_i$ of three different constructions of the logical CNOT gate indexed by $i$. However, the predictions and measurements of the construction \textbf{CNOT1f} (yellow), shows little dependence on the fitting parameter and was omitted from the fitting procedure. The other two constructions \textbf{CNOT2F} (green) and \textbf{CNOT3f} (purple) using the FT QEC, both of which are described in the text. Simulations of the experiments were run with a global coherent dephasing rate $\nu$ that was scanned over several different values from $0.2$ to $0.4$ Hz. The best fit parameter $\nu$ was determined as the value that minimized the value of $\sum_i(m_i-s_i(\nu))^2$. We show results associated with experiments using $\overline{Z}$ basis input states, but also carried out the procedure using the $\overline{X}$ and Bell basis input states. Similar procedures were carried out for the color code experiments.}
\label{fig_eff_dephasing_fqc_z}
\end{figure}

As mentioned in the main text, our simulations use a parameterized global coherent dephasing model. We used the experimental data reported in this work to tune this parameter so as to minimize the average discrepancy between the experimental and simulated data. A example of this minimization procedure is shown in Fig.~\ref{fig_eff_dephasing_fqc_z}. This procedure resulted in an effective dephasing rate of 0.236 Hz for the five-qubit code experiments in H1-2 and 0.172 Hz for the color code experiments in H1-1.

\begin{figure}[ht]
\includegraphics[width=\columnwidth]{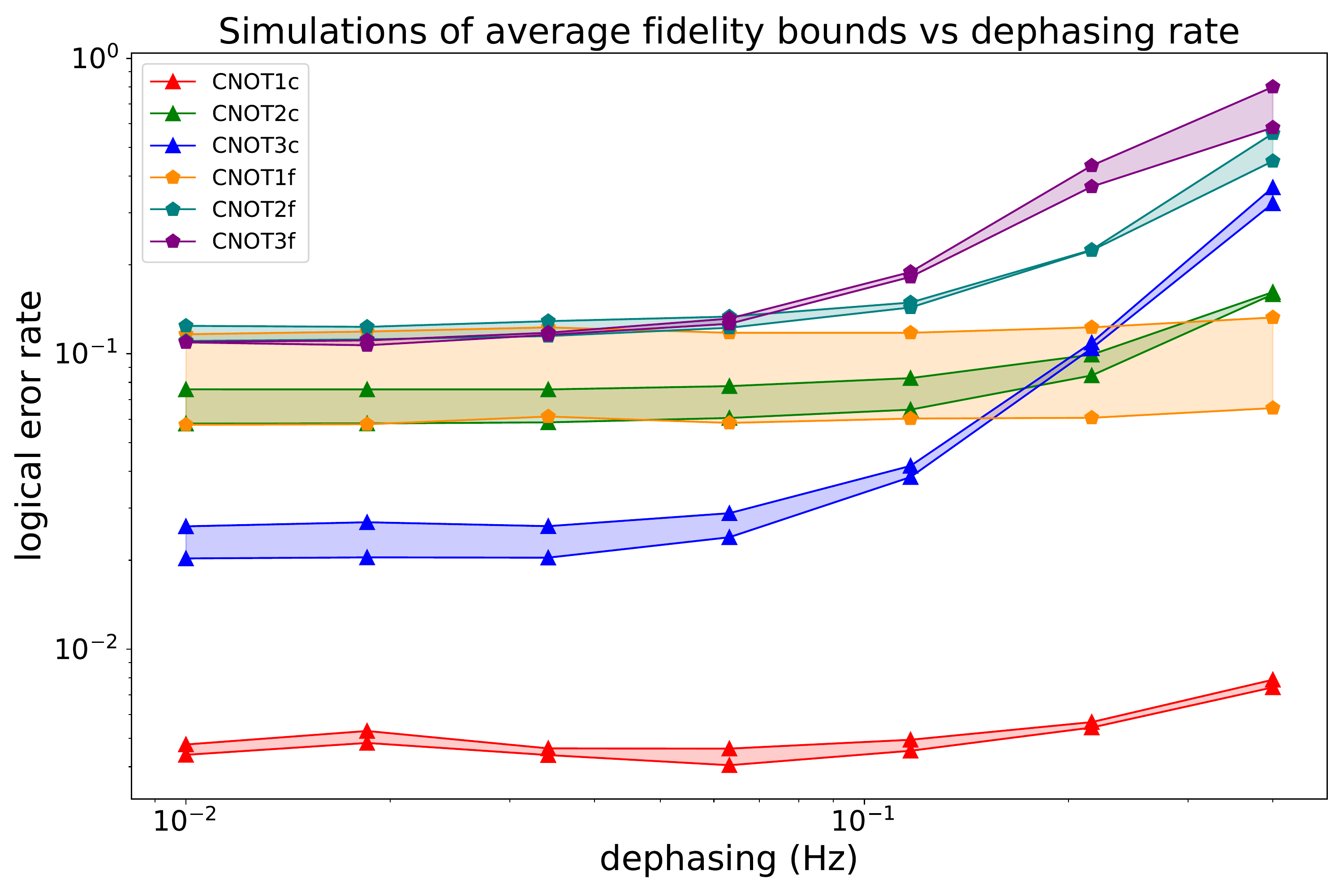}
\caption{Simulation results for five-qubit logical circuits using the $\overline{X}$ and $\overline{Z}$ basis states and Bell prep input states. The dephasing rate is scaled while keeping the rest of the error budget fixed.}
\label{fig_scale_dephasing_bounds}
\end{figure}

Here we present results from numerical simulations where the dephasing rate is scaled while keeping the rest of the error budget constant. These numerical experiments are motivated by the five-qubit code results which show deeper, more FT circuitry consistently degrades performance. Fig.~\ref{fig_scale_dephasing_bounds} shows that the average infidelity bound determined from simulating different circuits depends on the amount of dephasing in our model for both the five qubit code and the color code. From these bounds, we do not see that FT significantly improves the performance relative to the two non-FT experiments. However, there is some indication for small dephasing rates that the FT experiment \textbf{CNOT3f} might start to outperform the non-FT experiment \textbf{CNOT2f}, which utilizes a single round of syndrome extraction, but tighter bounds are required to state this definitively. Intriguingly, we see that with reduced dephasing the FT experiment with a QEC cycle (\textbf{CNOT2c}) outperforms the non-FT experiment with a single syndrome extraction round (\textbf{CNOT2c}).

\begin{figure}[ht]
\includegraphics[width=\columnwidth]{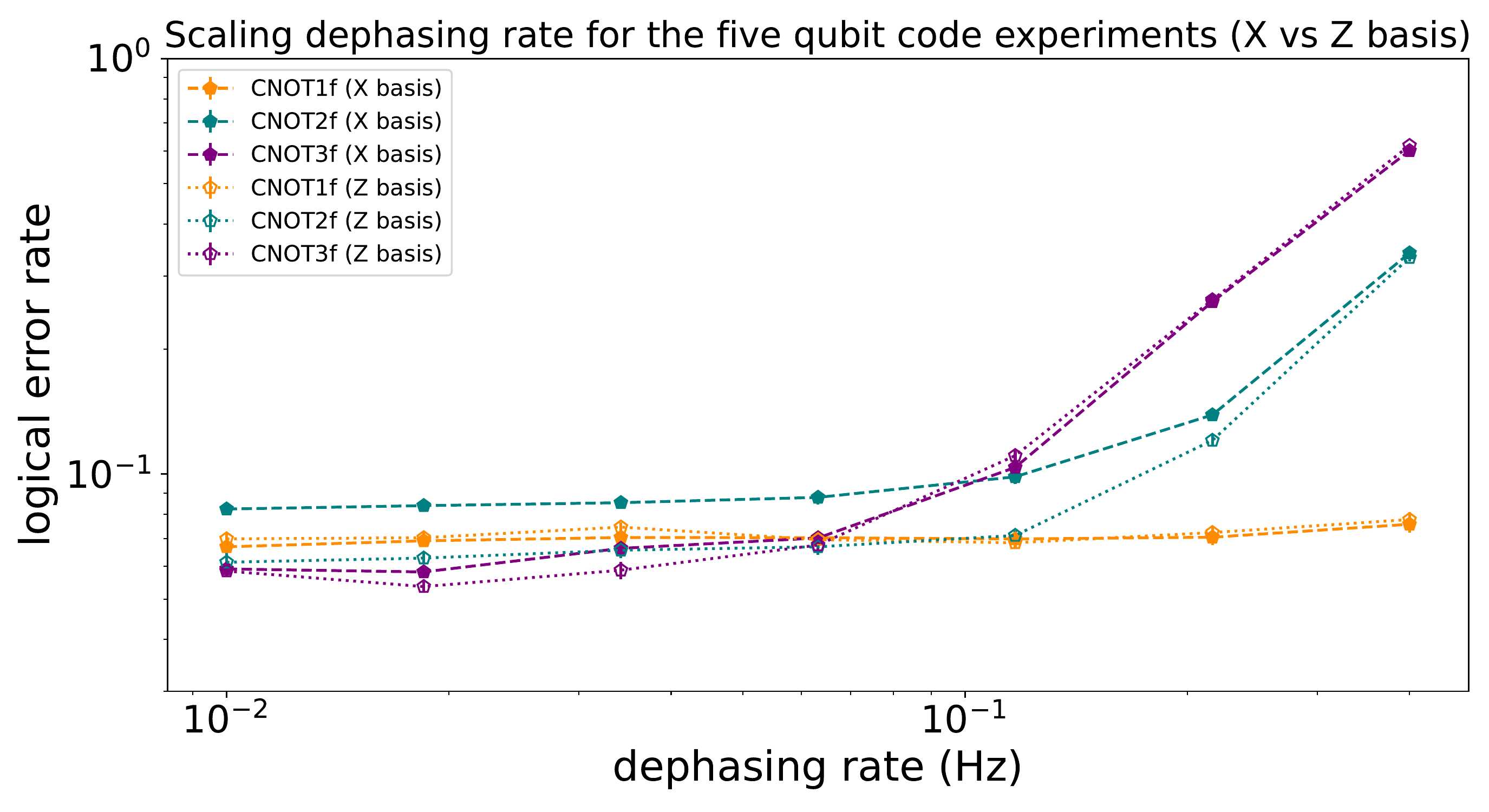}
\caption{Simulation results for five qubit code logical circuits using the $\overline{X}$ and $\overline{Z}$ basis input states. The dephasing rate is scaled while keeping the rest of the error budget fixed.}
\label{fig_scale_dephasing_fqc_xz}
\end{figure}

\begin{figure}[ht]
\includegraphics[width=\columnwidth]{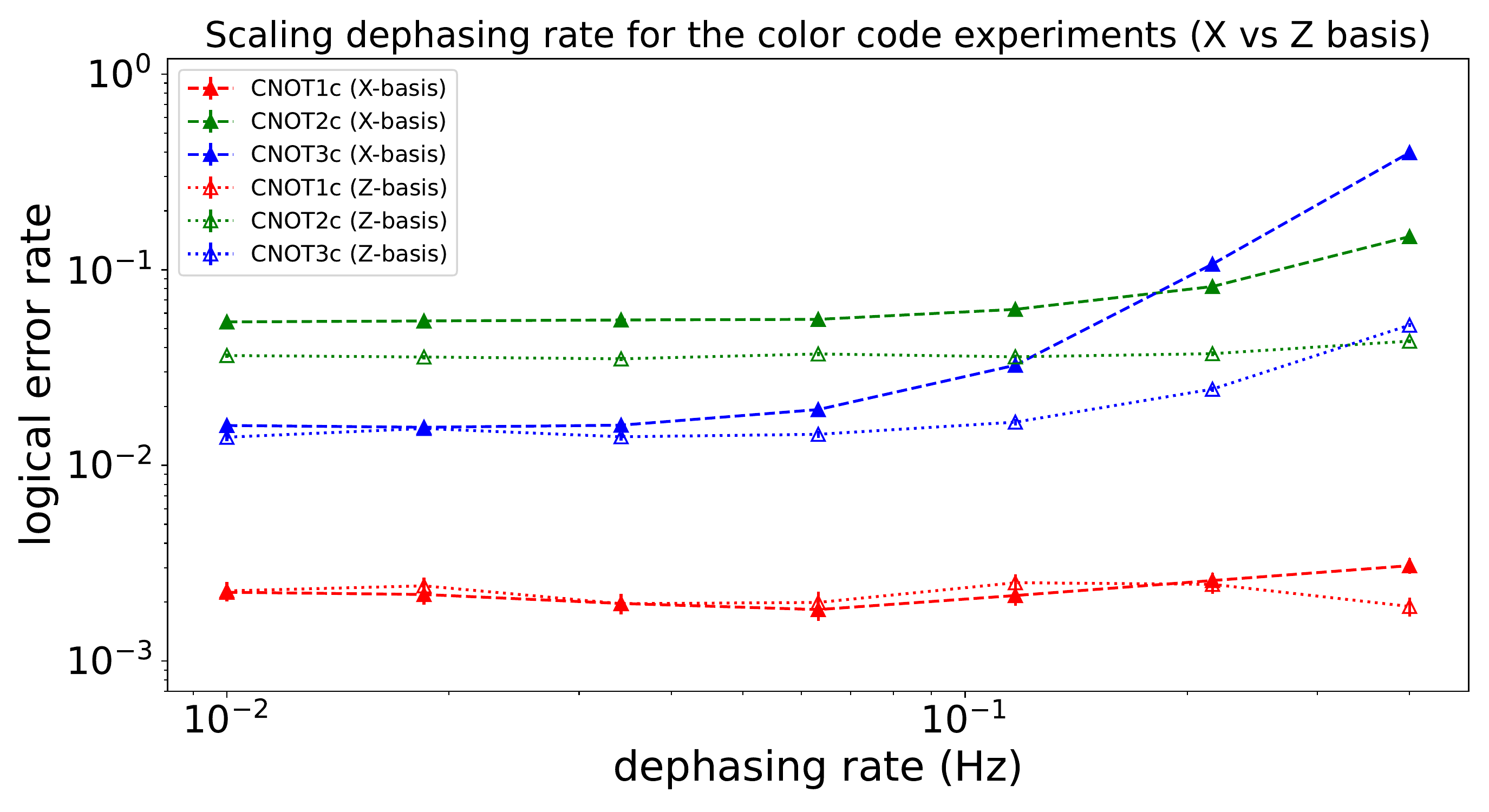}
\caption{Simulation results for color code logical circuits using the $\overline{X}$ and $\overline{Z}$ basis input states. The dephasing rate is scaled while keeping the rest of the error budget fixed.}
\label{fig_scale_dephasing_cc_xz}
\end{figure}

To evaluate the effect of bias Z noise due to dephasing, Fig.~\ref{fig_scale_dephasing_fqc_xz} and Fig.~\ref{fig_scale_dephasing_cc_xz} compares the logical error rates for $\overline{X}$ and $\overline{Z}$ basis states for the five-qubit and color code, respectively, to varying the dephasing rate. From this, five-qubit code appears to performs similarly in both the $\overline{X}$ and $\overline{Z}$ bases; however, the simulations predict the \textbf{CNOT2f} experiments, which has relatively high depth but is not FT, perform slightly better in the $\overline{X}$ compared to the low depth but not FT circuits of \textbf{CNOT1f} and the high depth but FT circuits of \textbf{CNOT3f}; therefore, this asymetric repose to bias noise maybe be due to a combination of larger depth, which accrues more dephasing noise, and lower corrective power due to utilizing a single non-flagging syndrome extraction round. For the color code, we see that deeper circuits, both FT and non-FT, we see a worse performance in the $\overline{Z}$, which is to be expected given the logical $Z$ operator is a weight three $Z$ operator and is consistant with the behavior seen in previous work ~\cite{RyanAnderson2021}. However, we do see that the experiment \textbf{CNOT1c} does not have a significant performance different between the two bases. This may be expected due to the small circuit depth leading to less build up of dephasing noise.  


\clearpage
\onecolumngrid
\section{Additional circuit diagrams}
Here we show additional details of the circuitry used in the five-qubit code experiments.
\begin{figure*}[ht!]
\includegraphics[trim=90 20 720 20, clip, scale=0.47]{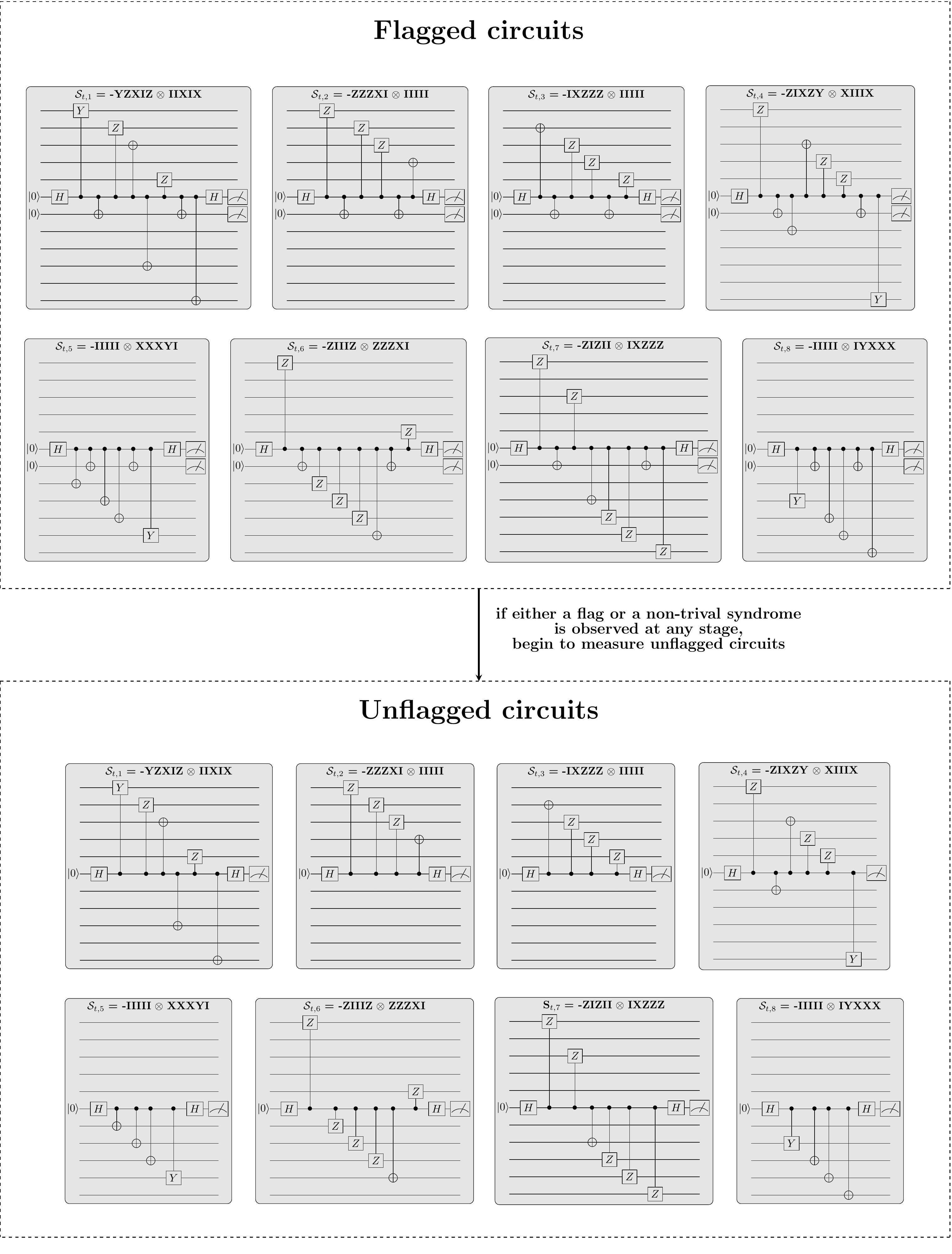}
\caption{Circuit diagrams showing how the stabilizers are measured in the two-qubit $[[10,2,3]]$ QEC cycle referenced in Fig.~\ref{main_char}. For clarity, the ancilla qubits are placed in the middle of the circuit, and the five qubits making up each code block are on the top and bottom and the full tensor product notation is dropped, except between the stabilizers of the individual code blocks. For the full FT QEC cycle (labeled \textbf{CNOT3f} in Table~\ref{FidelityTable_v1}), the flagged circuits are executed first. If either a non-trivial syndrome or a flag is raised, the circuit dynamically changes to measuring the unflagged checks. The resulting measurements from the flagged and unflagged checks provide enough syndrome information to decode FT. For the non-FT version only performing a single QEC cycle (labeled \textbf{CNOT2f} in Table \ref{FidelityTable_v1}), only the unflagged circuits are measured.}
\label{qec_cycle_flag}
\end{figure*}

\begin{figure*}[ht]
\includegraphics[trim=10 225 175 0, clip, width=\textwidth]{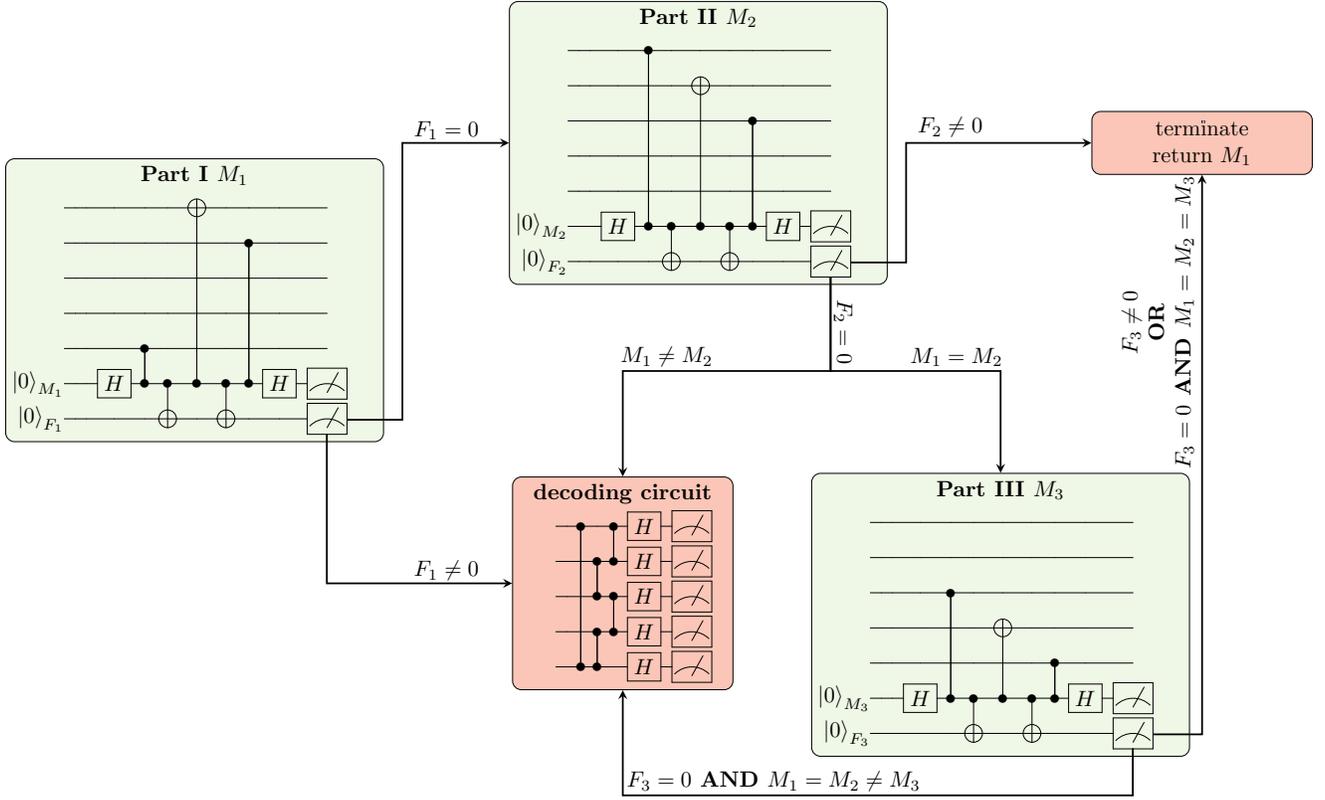}
\caption{An exploded view of the five-qubit code's adaptive measure out scheme referenced in the text and in Fig.~\ref{main_char}. The FT measure out circuit is similar to the reverse of the FT initialization circuit, but only part of the circuit is run in each trial, depending on the outcomes. The three green boxes labeled Part I, Part II and Part III are flagged measurements of $\overline{X}$ representatives with the measure and flagging qubits labeled $M_i$ and $F_i$. The three different weight-three logical representatives are related by multiplying by different stabilizers. If previous syndrome extraction circuits were ran, the measurement outcomes are multiplied by previous stabilizer outcomes to infer and compare logical representative measurements directly; therefore, $M_i$ represents the updated logical measurements. In Part I, if the flag is non-trivial, a fault was detected and we jump to the decoding circuit. In Part II, if a flag is non-trivial then a fault was detected, but we only need to be FT to a single fault and can assume Part I did not contain a fault and return $M_1$ as our logical measurement. If the flag is trivial, we have two sub-cases: (1) the logical measurements disagree and the fault could have been in Part I or Part II, so we jump to the decoding circuit, (2) both logical measurements agree and we continue to Part III. In Part III, if the flag is non-trivial a fault has been detected in this part and the result $M_1 = M_2$ is returned. If the flag was trivial, we again have two sub-cases: (1) if $M_1=M_2 \neq M_3$, we jump to the decoding circuit, (2) if $M_1=M_2=M_3$, we have not detected a fault and return this result. If we reach the decoding circuit, we reverse the encoding process and get a set of measurements that multiply together to infer the results of the four stabilizers and $\overline{X}$ operator. (We also mod out the results of previous stabilizer measurements made if previous syndrome extraction was preformed.) We then send the inferred stabilizer measurement to a classical decoder algorithm to determine a final logical Pauli correction, which is used to update the Pauli frame. Whether the logical measurement was determined through the decoding circuit or was returned at a previous stage, the logical measurement is corrected using the results of the stored Pauli frame.
}
\label{ft_meas}
\end{figure*}

\begin{figure}[ht]
\includegraphics[trim=0 340 575 0, clip, scale=0.9]{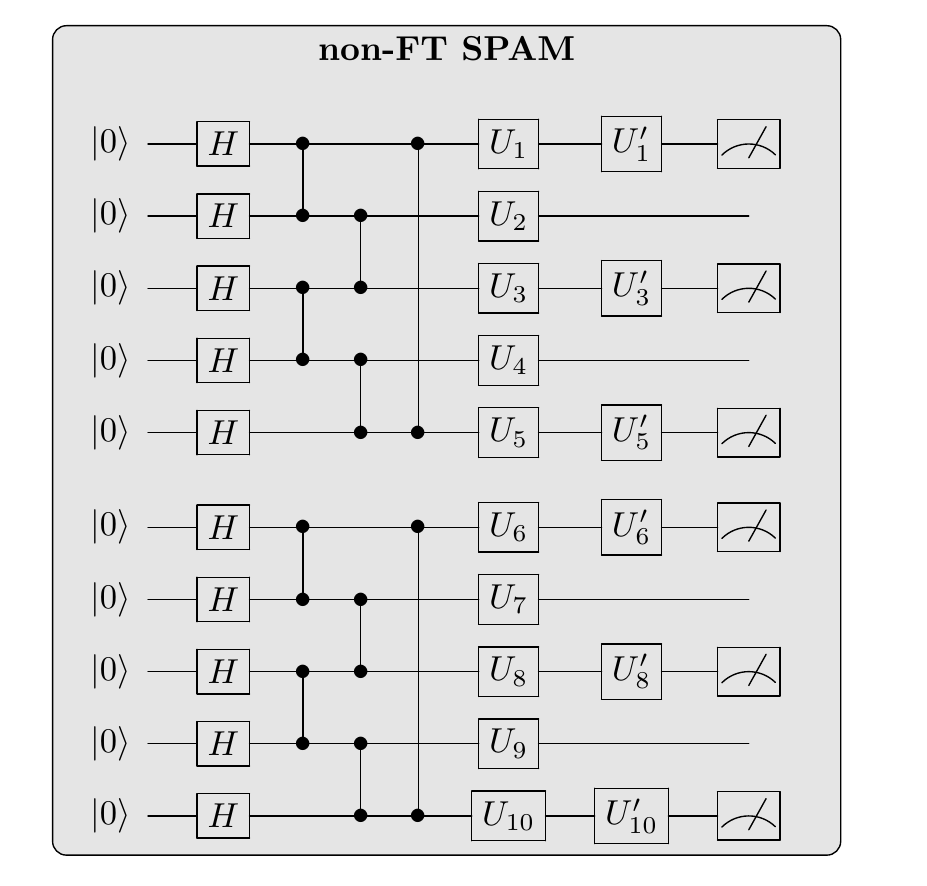}
\caption{The circuit used to measure the non-FT SPAM fidelity of the five-qubit code, the circuit is labeled \textbf{SPAM1f} in Table~\ref{FidelityTable_v1}. The variable unitaries are used to select a particular state to prepare and a particular measurement basis. With this non-FT measure out scheme, we can directly measure the raw logical output supported on qubit 1, 3, and 5.}
\label{nft_meas}
\end{figure}

\begin{figure*}[ht]
\includegraphics[trim=15 340 280 0, clip, scale=0.9]{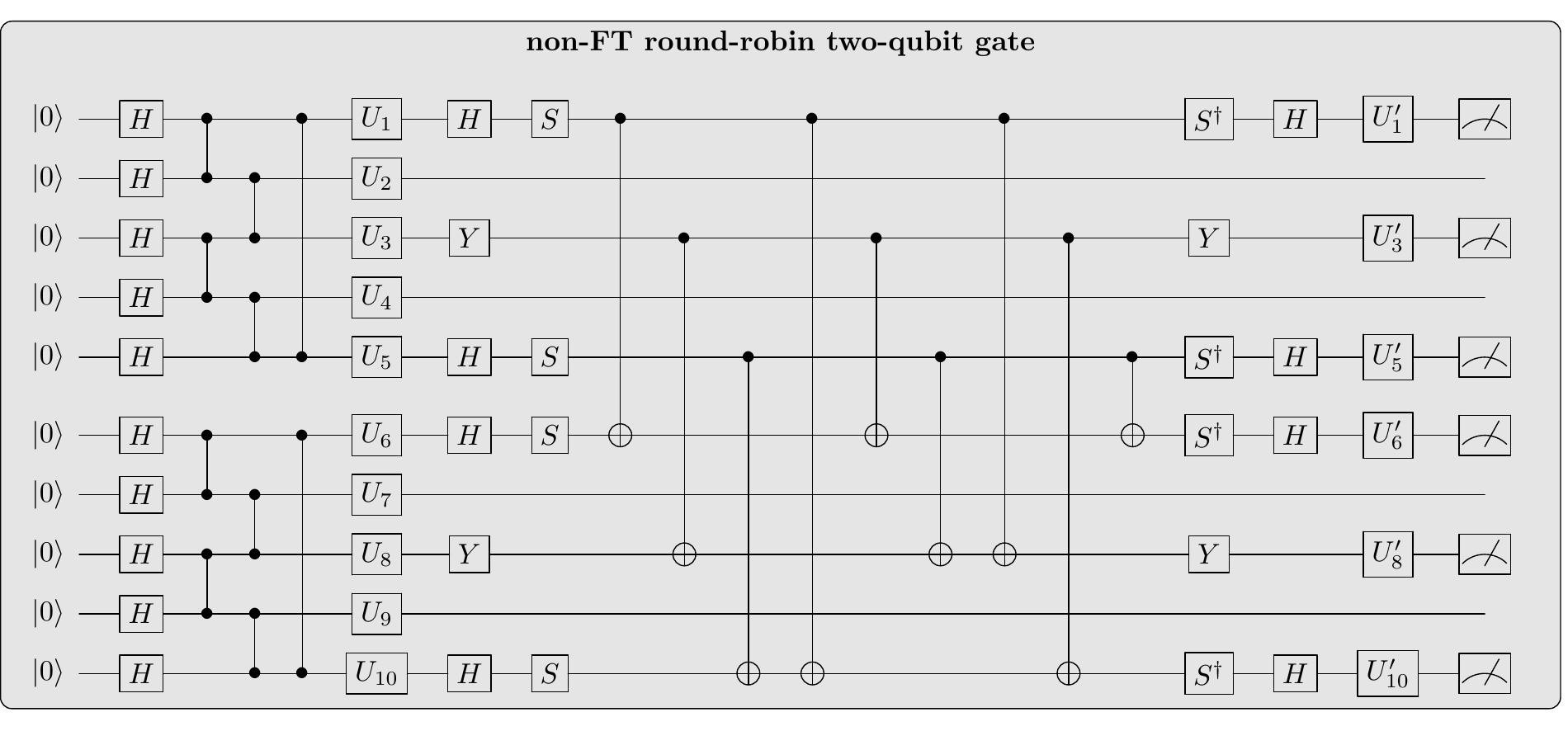}
\caption{The circuit used to implement the non-FT version of the five-qubit code round-robin gate, the circuit is labeled \textbf{CNOT1f} in Table~\ref{FidelityTable_v1}. The unprimed variable unitaries are used to select different input states, and the primed variable unitaries near the end of the circuit are used to select the measurement basis.}
\label{nft_TQG}
\end{figure*}

\end{document}